\documentclass[letterpaper]{article} 
\usepackage{aaai25}  
\usepackage{times}  
\usepackage{helvet}  
\usepackage{courier}  
\usepackage[hyphens]{url}  
\usepackage{graphicx} 
\usepackage{natbib}  
\usepackage{caption} 
\usepackage{algorithm}
\usepackage{algorithmic}

\usepackage{amsmath}
\usepackage{amssymb}
\usepackage{mathtools}
\usepackage{amsthm}
\usepackage{cuted}

\theoremstyle{plain}
\newtheorem{theorem}{Theorem}

\newtheorem{lemma}[theorem]{Lemma}

\theoremstyle{definition}

\theoremstyle{remark}

\usepackage[textsize=tiny]{todonotes}

\usepackage{url}
\usepackage[para,online,flushleft]{threeparttable}

\usepackage[
  font = small,
  labelfont = bf,
  tableposition = top
]{caption}
\usepackage{subcaption}

\usepackage{booktabs}

\usepackage{mathrsfs}
\usepackage{physics}

\usepackage{comment}
\usepackage{enumitem, multirow, xspace}

\usepackage[]{color-edits}
\allowdisplaybreaks[4]

\usepackage{newfloat}
\usepackage{listings}
\DeclareCaptionStyle{ruled}{labelfont=normalfont,labelsep=colon,strut=off} 
\newcommand{\name}{NASM\xspace}

\title{Optimal Control Operator Perspective and a Neural Adaptive Spectral Method}
\author{
    Mingquan Feng\textsuperscript{\rm 1},
    Zhijie Chen\textsuperscript{\rm 2},
    Yixin Huang\textsuperscript{\rm 1},
    Yizhou Liu\textsuperscript{\rm 1},
    Junchi Yan\textsuperscript{\rm 1,3\footnote{Corresponding author. This work was in part supported by the NSFC (62222607, 623B1009).}}
}
\affiliations{
    
    \textsuperscript{\rm 1}School of Electronic Information \& Electrical Engineering, 
Shanghai Jiao Tong University, Shanghai, China\\
\textsuperscript{\rm 2}Siebel School of Computing and Data Science, University of Illinois at Urbana-Champaign, Urbana, IL, USA \\
    \textsuperscript{\rm 3}School of Artificial Intelligence \& MoE Lab of AI, Shanghai Jiao Tong University, Shanghai, China\\
    \{fengmingquan, cindyh1103, yizhou0409, yanjunchi\}@sjtu.edu.cn, lucmon@illinois.edu
}

\begin{document}

\maketitle

\begin{abstract}
Optimal control problems (OCPs) involve finding a control function for a dynamical system such that a cost functional is optimized. It is central to physical systems in both academia and industry. In this paper, we propose a novel instance-solution control operator perspective, which solves OCPs in a one-shot manner without direct dependence on the explicit expression of dynamics or iterative optimization processes. The control operator is implemented by a new neural operator architecture named Neural Adaptive Spectral Method (NASM), a generalization of classical spectral methods. We theoretically validate the perspective and architecture by presenting the approximation error bounds of NASM for the control operator.  Experiments on synthetic environments and a real-world dataset verify the effectiveness and efficiency of our approach, including substantial speedup in running time, and high-quality in- and out-of-distribution generalization. 
\end{abstract}

%
\begin{links}
     \link{Code}{https://github.com/FengMingquan-sjtu/NASM.git}
\end{links}

\section{Introduction}
Although control theory has been rooted in a model-based design and solving paradigm, the demands of model reusability and the opacity of complex dynamical systems call for a rapprochement of modern control theory, machine learning, and optimization. Recent years have witnessed the emerging trends of control theories with successful applications to engineering and scientific research, such as robotics \cite{krimsky2020optimal}, aerospace technology \cite{he2019robust}, and economics and management \cite{lapin2019numerical} etc.

We consider the well-established formulation of optimal control~\citep{kirk2004optimal} in finite time horizon $T=[t_0, t_f]$. Denote $X$ and $U$ as two vector-valued function sets, representing state functions and control functions respectively. Functions in $X$ (resp. $U$) are defined over $T$ and have their outputs in $\mathbb{R}^{d_x}$ (resp. $\mathbb{R}^{d_u}$). State functions $\vb*{x} \in  X$ and control functions $\vb*{u} \in U$ are governed by a differential equation. The optimal control problem (OCP) is targeted at finding a control function that minimizes the cost functional $f$ \citep{kirk2004optimal,lewis2012optimal}:
\begin{subequations}
\begin{align}
\min_{\vb*{u} \in U} && f(\vb*{x}, \vb*{u}) &= \int_{t_0}^{t_f} p(\vb*{x}(t), \vb*{u}(t)) \dd{t} + h(\vb*{x}(t_f))\\
\text{s.t.}&&\dot{\vb*{x}}(t) &= \vb*{d}(\vb*{x}(t), \vb*{u}(t)),  \\
&& \vb*{x}(t_0) &= \vb*{x}_{0},
\end{align}
\label{eq:oc}
\end{subequations} 
where $\vb*{d}$ is the dynamics of differential equations; $p$ evaluates the cost alongside the dynamics and $h$ evaluates the cost at the termination state $\vb*{x}(t_f)$;  and $\vb*{x}_{0}$ is the initial state. We restrict our discussion to differential equation-governed optimal control problems, leaving the control problems in stochastic networks~\citep{dai2022queueing}, inventory management~\citep{abdolazimi2021designing}, etc. out of the scope of this paper. The analytic solution of Eq.~\ref{eq:oc} is usually unavailable, especially for complex dynamical systems. Thus, there has been a wealth of research towards accurate, efficient, and scalable numerical OCP solvers \citep{rao2009survey} and neural network-based solvers \citep{kiumarsi2017optimal}. However, both classic and modern OCP solvers are facing challenges, especially in the big data era, as briefly discussed below.

\textbf{1) Opacity of Dynamical Systems.} Existing works~\citep{Böhme2017direct,effati2013optimal, jin2020pontryagin} assume the dynamical systems a priori and exploit their explicit forms to ease the optimization. However, real-world dynamical systems can be unknown and hard to model. It raises serious challenges in data collection and system identification~\citep{ghosh2021variational}, where special care is required for error reduction.

\textbf{2) Model Reusability.} Model reusability is conceptually measured by the capability of utilizing historical data when facing an unprecedented problem instance. Since solving an individual instance of Eq.~\ref{eq:oc} from scratch is expensive, a reusable model that can be well adapted to new problems is welcomed for practical usage. 


\textbf{3) Running Paradigm.} Numerical optimal control solvers traditionally use iterative methods before picking the control solution, thus introducing a multiplicative term regarding the iteration in the running time complexity. This sheds light on the high cost of solving a single OC problem.

\textbf{4) Control Solution Continuity.} Control functions are defined on a continuous domain (typically time) despite being intractable for numerical solvers. Hence resorting to discretization in the input domain gives rise to the trade-off in the precision of the control solution and the computational cost, as well as the truncation errors. While the discrete solution can give point-wise queries, learning a control function for arbitrary time queries is much more valued.

\begin{figure}[tb!]
\centering
\includegraphics[width=0.96\linewidth]{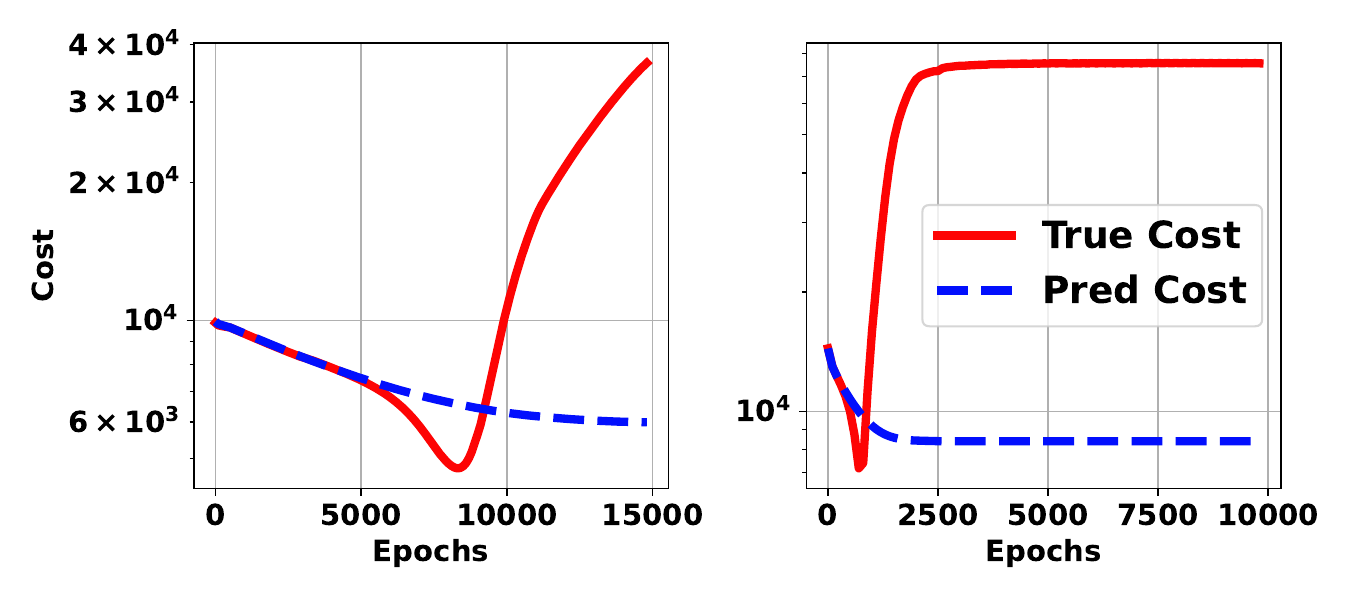}
    \caption{Phase-2 cost curves of two failed instances of two-phase control \citep{hwang2022solving} on Pendulum system. The control function gradually moves outside the training  distribution of phase 1. As a result, the control function converges w.r.t. the cost predicted by the surrogate model (blue), but diverges w.r.t. true cost (red).}
    \label{fig:PendulumTP}
\end{figure}

\textbf{5) Running Phase.} The two-phase models~\citep{chen2018optimal, wang2021fast,hwang2022solving} can (to some extent) overcome the above issues at the cost of introducing an auxiliary dynamics inference phase. This thread of works first approximate the state dynamics by a differentiable surrogate model and then, in its second phase, solves an optimization problem for control variables. However, the two-phase paradigm increases computational cost and manifests inconsistency between the two phases. A motivating example in Fig.~\ref{fig:PendulumTP} shows that the two-phase paradigm leads to failures. When the domain of phase-2 optimization goes outside the training distribution in phase-1, it might collapse.

Table~\ref{table:1} compares the methods regarding the above aspects. We propose an instance-solution operator perspective for learning to solve OCPs, thereby tackling the issues above. \textbf{The highlights are:}
%
\begin{table*}[!tb]
\caption{Optimal control approaches. Our NCO covers all the merits of performing a single-phase direct-mapping paradigm without relying on known system dynamics and supports arbitrary input-domain queries.}
\label{tab:methodcompare}
\begin{center}
\begin{threeparttable}
\begin{tabular}{l c c c c c}
    \toprule
    \textbf{Methods} & \textbf{Phase} & \textbf{Continuity} & \textbf{Dynamics} & \textbf{Reusability} & \textbf{Paradigm} \\ \midrule
      Direct~\cite{Böhme2017direct}   &  Single   &  Discrete  &  Required   &  No  &  Iterative \\
     Two-Phase~\cite{hwang2022solving}   & Two   &  Discrete   & Dispensable\tnote{1} & Partial\tnote{2} &  Iterative \\ 
     PDP~\cite{jin2020pontryagin}  & Single  &  Discrete  &  Required   &  No  &  Iterative  \\
     DP~\citep{tassa2014control}  &  Single  & Discrete  & Required &   No                 &  Iterative \\ \midrule
      \textbf{NCO (ours)} &  Single  &  Continuous  & Dispensable  &  Yes  &  One-pass\\
     \bottomrule
\end{tabular}
\begin{tablenotes}
\item[1] if phase-1 uses PINN loss \citep{wang2021fast}: required. \item[2] only phase-1 is reusable.
\end{tablenotes}
\end{threeparttable}
\end{center}
\label{table:1}
\end{table*}

    
    \textbf{1)} We propose the Neural Control Operator (NCO) method of optimal control, solving OCPs by learning a neural operator that directly maps problem instances to solutions. The system dynamics is implicitly learned during the training, which relies on neither any explicit form of the system nor the optimization process at test time. The operator can be reused for similar OCPs from the same distribution without retraining, unlike learning-free solvers. Furthermore, the single-phase direct mapping paradigm avoids iterative processes with substantial speedup. 

     \textbf{2)} We instantiate a novel neural operator architecture: NASM (Neural Adaptive Spectral Method) to implement the NCO and derive bounds on its approximation error.
   
     \textbf{3)} Experiments on both synthetic and real systems show that \name is versatile for various forms of OCPs. It shows over 6Kx speedup (on synthetic environments) over classical direct methods. It also generalizes well on both in- and out-of-distribution OCP instances, outperforming other neural operator architectures in most of the experiments.

\textbf{Background and Related Works.} Most OCPs can not be solved analytically, thus numerical methods \citep{korkel2004numerical} are developed. An OCP numerical method has three components: problem formulation, discretization scheme, and solver. The problem formulations fall into three categories: direct methods, indirect methods, and dynamic programming. Each converts OCPs into infinite-dimensional optimizations or differential equations, which are then discretized into finite-dimensional problems using schemes like finite difference, finite element, or spectral methods. These finite problems are solved by algorithms such as interior-point methods for nonlinear programming and the shooting method for boundary-value problems. These solvers rely on costly iterative algorithms and have limited ability to reuse historical data.

In addition, our work is also related to the neural operators, which are originally designed for solving parametric PDEs. Those neural operators learn a mapping from the parameters of a PDE to its solution. For example, DeepONet (DON)~\citep{lu2021learning} consists of two sub-networks: a branch net for input functions (i.e. parameters of PDE) and a trunk net for the querying locations or time indices. The output is obtained by computing the inner product of the outputs of branch and trunk networks. Another type of neural operator is Fourier transform based, such as Fourier Neural Operator (FNO)~\citep{li2020fourier}. The FNO learns parametric linear functions in the frequency domain, along with nonlinear functions in the time domain. The conversion between those two domains is realized by discrete Fourier transformation. There are many more architectures, such as Graph Element Network (GEN)~\citep{alet2019graph}, Spectral Neural Operator(SNO)~\citep{fanaskov2022spectral} etc., as will be compared in our experiments.


\section{Methodology}
In this section, we will present the instance-solution control operator perspective for solving OCPs.  The input of operator $\displaystyle \mathcal{G}$ is an OCP instance $i$, and the output is the optimal control function $\vb*{u}^\ast$, i.e. $\displaystyle \mathcal{G}:I \rightarrow U$.  Then we propose the Neural Control Operator (NCO), a class of end-to-end OCP neural solvers that learn the underlying infinite-dimensional operator $\displaystyle \mathcal{G}$ via a neural operator. A novel neural operator architecture, the Neural Adaptive Spectral Method (NASM), is implemented for NCO method.

\subsection{Instance-Solution Control Operator Perspective of OCP Solvers}
This section presents a high-level instance-solution control operator perspective of OCPs. In this perspective, an OCP solver is an operator that maps problem instances (defined by cost functionals, dynamics, and initial conditions) into their solutions, i.e. the optimal controls. Denote the problem instance by $i = (f, \vb*{d}, \vb*{x}_{init}) \in I$, with its optimal control solution $\vb*{u}^* \in U$. The operator is defined as $\mathcal{G}:I \rightarrow U$, a mapping from cost functional to optimal control.

In practice, the non-linear operator $\mathcal{G}$ can hardly have a closed-form expression. Therefore, various numerical solvers have been developed, such as direct method and indirect method, etc, as enumerated in the related works. All those numerical solvers can be regarded as approximate operators $\mathcal{N}:I \rightarrow U$ of the exact operator $\mathcal{G}$. To measure the quality of the approximated operator, we define the approximation error as the distance between the cost of the solution and the optimal cost. Let $\mu$ be the probability measure of the problem instances, assume the control dimension $d_{u}=1$, and cost $f_i>0$ (subscript $i$ means the cost is dependent on $i$), w.l.o.g. Then the \emph{approximation error} is defined as:
\begin{equation}
    \widehat{\mathscr{E}} := \int_{I} \abs{ \frac{f_{i} \circ \mathcal{N}(i) - f_{i} \circ \mathcal{G}(i)}{f_{i} \circ \mathcal{G}(i)} } \dd \mu(i).
\label{eq:approx err}
\end{equation}

Classical numerical solvers are generally guaranteed to achieve low approximation error (with intensive computing cost). For efficiency, we propose to \emph{approximate the operator $\mathcal{G}$ by neural networks}, i.e. \textbf{Neural Control Operator (NCO)} method. The networks are trained by demonstration data pairs $(i, \vb*{u}^*)$ produced by some numerical solvers, without the need for explicit knowledge of dynamics.  Once trained, they infer optimal control for any unseen instances with a single forward pass. As empirically shown in our experiments, the efficiency (at inference) of neural operators is consistently better than classical solvers. We will also provide an approximation error bound for our specific implementation NASM of NCO.

Before introducing the network architecture, it is necessary to clarify a common component, the Encoder, used in both classical and neural control operators. The encoder converts the infinite-dimensional input $i$ to finite-dimensional representation $\vb*{e}$. Take the cost functional for example. If the cost functional $f(x,u)=\int (x-x_\text{target})^2 \dd t$ is explicitly known, then the $f$ may be encoded as the symbolic expression or the parameters $x_\text{target}$ only. Otherwise, the cost functional can be represented by sampling on grids. After encoding, The encoded vector $\vb*{e}$ is fed into numerical algorithms or neural networks in succeeding modules. The encoder's design is typically driven by the problem setting rather than the specific solver or network architecture.

\subsection{Neural Adaptive Spectral Method}
\label{sec:arch}
From the operator's perspective, constructing an OCP solver is equivalent to an operator-learning problem. We propose to learn the operator by neural network in a data-driven manner. Now we elaborate on a novel neural operator architecture named Neural Adaptive Spectral Method (NASM), inspired by the spectral method. 
  
The spectral method assumes that the solution is the linear combination of $p$ basis functions:
  \begin{equation}
  \label{eq: spectral method}
      \mathcal{N}_\text{spec}(i)(t) = \sum_{j=1}^{p} c_j(i) b_j(t),
  \end{equation}
    where the basis functions $\{b_j\}_p$ are chosen as orthogonal polynomials e.g. Fourier series (trigonometric polynomial) and Chebyshev polynomials. The coefficients $\{c_j\}_p$ are derived from instance $i$ by a numerical algorithm. The Spectral Neural Operator (SNO)~\cite{fanaskov2022spectral} obtains the coefficients by a network.

Our NASM model is based on a similar idea but with more flexible and adaptive components than SNO. The first difference is that summation $\sum$ is extended to aggregation $\bigoplus$. The aggregation is defined as any mapping $\mathbb{R}^p \rightarrow \mathbb{R}$ and can be implemented by either summation or another network. Secondly, the coefficient $\{c_j\}_p$ is obtained from Coefficient Net, which is dependent on not only instance $i$ but time index $t$ as well, inspired by time-frequency analysis \citep{cohen1995time}. It is a generalization and refinement of $\mathcal{N}_\text{spec}$(Eq.~\ref{eq: spectral method}), for the case when the characteristics of $\mathcal{G}$ are non-static.  For example, $\mathcal{N}_\text{spec}$ with Fourier basis can only capture globally periodic functions, while the $\mathcal{N}_\text{NASM}$ with the same basis can model both periodic and non-periodic functions. Lastly, the basis functions $\{b_j\}_p$ are adaptive instead of fixed. The adaptiveness is realized by parameterizing basis functions by $\theta$, which is inferred from $t,i$ also by the neural network.

\begin{equation}
      \mathcal{N}_\text{NASM}(i)(t) = \bigoplus_{j=1}^{p} c_j(t,i) b_j(t; \vb*{\theta}(t,i)).
  \end{equation}
  
As a concrete example of adaptive basis functions, the adaptive Fourier series is obtained from the original basis by scaling and shifting, according to the parameter $\vb*{\theta}$. We limit the absolute value of elements of $\vb*{\theta}$ to avoid overlapping the adaptive range of basis functions. Following this principle, one can design adaptive versions for other basis function sets.
\begin{equation}
\begin{aligned}
    \{b_j(t; \vb*{\theta})\} = \{1, \:\: \sin(\pi [(1+\theta_{1}) t + \theta_{2}]), \\ 
    \:\: \cos(\pi [(1+\theta_{3}) t + \theta_{4}]), \:\: \cdots \}, \quad \abs{\theta_k} \le 0.5, \forall k.
\end{aligned}
\end{equation}

\begin{figure*}[t!]
    \centering
    \includegraphics[width=0.86\linewidth]{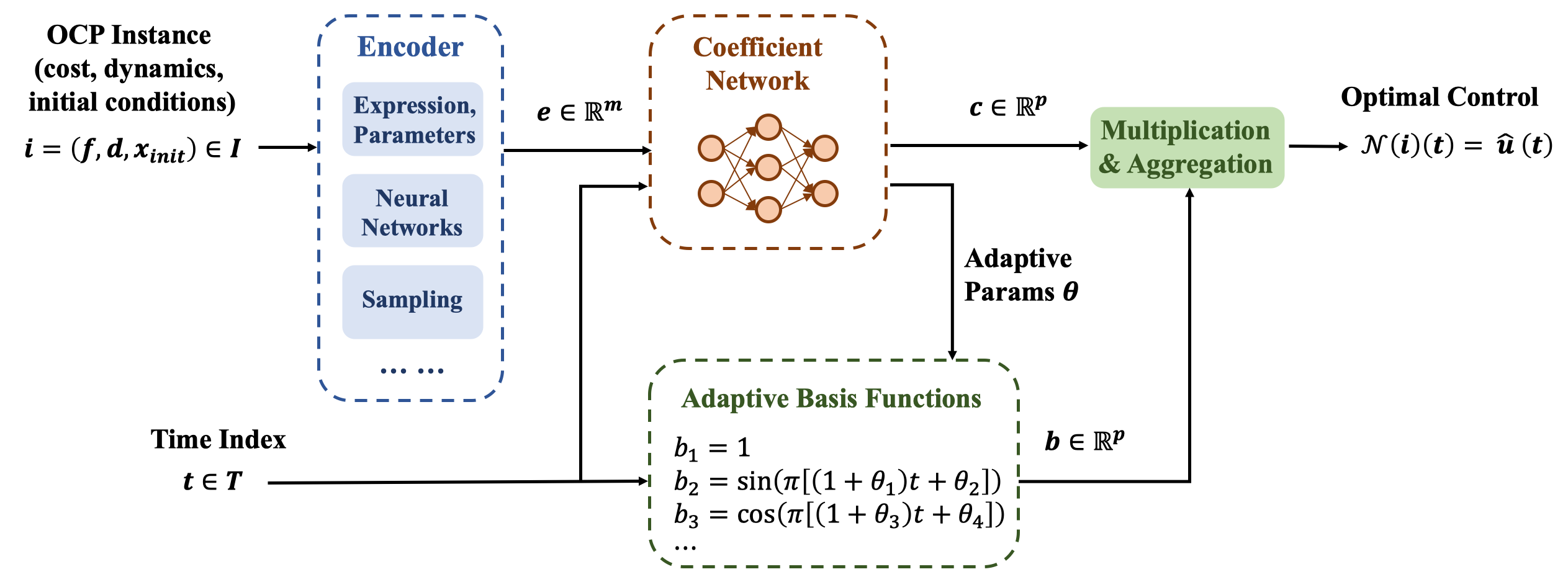}
    \caption{ The architecture of NASM. The network takes two inputs: OCP instance $i$ and time index $t$. The input $i$ is pre-processed by the Encoder. Then both $t$ and encoding $\vb*{e}$ are fed into the Coefficient Network to obtain coefficients $\vb*{c}$ and adaptive parameters $\vb*{\theta}$. The adaptive basis (e.g. Fourier series) outputs function values $\vb*{b}$, which is multiplied with $\vb*{c}$ and aggregated to the final output $\hat{\vb*{u}}(t)$, the estimation of optimal control for instance $i$ at time $t$. Detailed explanation is given in Section~\ref{sec:arch}.}
    \label{fig:model}
\end{figure*}

  The overall architecture of NASM is given in Fig. \ref{fig:model}. The theoretic result guarantees that there exists a network instance of \name approximating the operator $\mathcal{G}$ to arbitrary error tolerance. Furthermore, the size and depth of such a network are upper-bounded. The technical line of our analysis is partly inspired by \citep{lanthaler2022error} providing error estimation for DeepONets. 
  \begin{theorem}[\textbf{Approximator Error, Informal}] Under some regularity assumptions, given an operator $\mathcal{G}$, and any error budget $\epsilon$, there exists a NASM satisfying the budget with bounded network size and depth.
  \end{theorem}
  

    Comparing NASM error bound with that of DeepONet, we discover that NASM can achieve the same error bound as DeepONet with fewer learnable parameters. More details of theorems and proofs are presented in Appx.\ref{appx: appx err thoerem} and Appx.\ref{appx:proofs}.

\section{Experiments}
We give experiments on synthetic and real-world environments to evaluate our instance-solution operator framework.
\subsection{Synthetic Control Systems}
\label{sec:synthetic exp}
\subsubsection{Control Systems and Data Generation}
We evaluate \name on five representative optimal control systems by following the same protocol of \citep{jin2020pontryagin}, as summarized in Table \ref{tab:systems}. We postpone the rest systems to Appendix \ref{appx: Dynamic Systems}, and only describe the details of the Quadrotor control here:
\begin{equation*}
\begin{aligned}
    \dot{\vb*{p}} = \vb*{v}, \qquad
    m\dot{\vb*{v}} = \bmqty{0 \\ 0 \\ mg} + \vb*{R}^\top(\vb*{q})\bmqty{0 \\ 0 \\ \vb*{1}^\top \vb*{u}} ,\\
    \dot{\vb*{q}} = \frac{1}{2} \vb*{\Omega}(\vb*{\omega}) \vb*{q},  \qquad
    \vb*{J} \dot{\vb*{\omega}} = \vb*{T}\vb*{u} - \vb*{\omega} \times \vb*{J} \vb*{\omega} .
\end{aligned}
\end{equation*}

This system describes the dynamics of a helicopter with four rotors. The state $\vb*{x} = [\vb*{p}^\top, \vb*{v}^\top, \vb*{\omega}^\top]^\top \in \mathbb{R}^9$ consists of parts: position $\vb*{p}$, velocity $\vb*{v}$, and angular velocity $\vb*{\omega}$. The control $\vb*{u}\in \mathbb{R}^4$ is the thrusts of the four rotating propellers of the quadrotor. $\vb*{q} \in \mathbb{R}^4$ is the unit quaternion \citep{jia2019quaternions} representing the attitude (spacial rotation) of quadrotor w.r.t. the inertial frame. $\vb*{J}$ is the moment of inertia in the quadrotor's frame, and $\vb*{T}\vb*{u}$ is the torque applied to the quadrotor.  We set the initial state $\vb*{x}_{init}=[[-8,-6,9]^\top, \vb*{0}, \vb*{0}]^\top$, the initial quaternion $\vb*{q}_{init} = \vb*{0}$.  The matrices $\vb*{\Omega}(\vb*{\omega}), \vb*{R}(\vb*{q}), \vb*{T}$ are coefficient matrices, see definition in Appx.~\ref{appx:quadrotor}. The cost functional is defined as $\int_0^{tf} \vb*{c_x}^\top(\vb*{x}(t)-\vb*{x}_{goal})^2  + c_u \norm{\vb*{u}(t)}^2  \dd{t}$, with coefficients $\vb*{c_x} = \vb*{1}$, $c_u=0.1$.

In experiments presented in this section, unless otherwise specified, we temporally fix the cost functional symbolic expression, dynamics parameters, and initial condition in both training and testing. In this setting, the solution of OCP only depends on the parameter of cost, i.e. target state $\vb*{x}_{goal}$. The input of NCO is the cost functional only, and the information of dynamics, and initial conditions are explicitly learned. Therefore, we generate datasets (for model training/validation) and benchmarks (for model testing) by sampling target states from a pre-defined distribution. To fully evaluate the generalization ability, we define both in-distribution (ID) and out-of-distribution (OOD) \citep{shen2021towards}. Specifically, we design two random variables, $\displaystyle \vb{x}^{in}_{goal} := \vb*{x}_{goal}^{base} + \vb*{\epsilon}_{in}$, and $\displaystyle \vb{x}^{out}_{goal} := \vb*{x}_{goal}^{base} + \vb*{\epsilon}_{out}$, where $\vb*{x}_{goal}^{base}$ is a baseline goal state, and $\vb*{\epsilon}_{in, out}$ are different noise applied to ID and OOD. In Quadrotor problems for example, we set $\vb*{x}_{goal}^{base}=\vb*{0.6}$, and uniform noise $\vb*{\epsilon}_{in} \sim \mathcal{U}(-0.5, 0.5) $, and $\vb*{\epsilon}_{out} \sim \mathcal{U}(-0.7, -0.5)$.  

The training data are sampled from ID, while validation data and benchmark sets are both sampled from ID and OOD separately. The data generation process is shown in Alg. \ref{alg:data gen} in Appendix. For a given distribution, we sample a target state $\vb*{x}_{goal}$ and construct the corresponding cost functional $f$ and OCP instance $i$. Then define 100 time indices uniformly spaced in time horizon $T=[0,\text{tf}]$, $\text{tf} \sim \mathcal{U}(1, 1.01)$. The length of $T$ is slightly perturbed to add noise to the dataset. Then we solve the resulting OCP by the Direct Method(DM) solver and get the optimal control $u^*$ at those time indices. After that we sample 10 time indices $\{t_j\}_{1\le j \le 10}$, creating 10 triplets $\{(f, t_j, u^*(t_j))\}_{1\le j \le 10}$ and adding them to the dataset. Repeat the process, until the size meets the requirement. The benchmark set is generated in the same way, but we store $(f, J_{opt})$ pairs ($J_{opt}$ is the optimal cost) instead of the optimal control.  The dataset with all three factors (cost functional, dynamics parameters, and initial condition) changeable can be generated similarly. 

\subsubsection{Implementation and Baselines}
 For all systems and all neural models, the loss is the mean squared error defined below, with a dataset $D$ of $N$ samples:
$
 	\mathcal{L} = \frac{1}{N} \sum_{i,j \in D} \norm{\mathcal{N}(i)(t_j) - \vb*{u}_i^*(t_j)}^2.
$, the learning rate starts from 0.01, decaying every 1,000 epochs at a rate of 0.9. The batch size is $\min(10,000,N)$, and the optimizer is Adam~\citep{kingma2014adam}. The NASM uses 11 basis functions of the Fourier series.  For comparison, we choose the following baselines. Other details of implementation and baselines are recorded in Appendix~\ref{appx: details on exp}.

1) \textbf{Direct Method (DM):} a classical direct OCP solver, with finite difference discretization and Interior Point OPTimizer (IPOPT) \citep{biegler2009large} backend NLP solver.

2) \textbf{Pontryagin Differentiable Programming (PDP)} \citep{jin2020pontryagin}: an adjoint-based indirect method,  differentiating through PMP, and optimized by gradient descent.

3) \textbf{DeepONet (DON)} \citep{lu2021learning}: The seminal work of neural operator. The DON output is a linear combination of basis function values, and both coefficients and basis are pure neural networks.

4) \textbf{Multi-layer Perceptron (MLP)}: A fully connected network implementation of the neural operator.

5) \textbf{Fourier Neural Operator (FNO)} \citep{li2020fourier}: A neural operator with consecutive Fourier transform layers, and fully connected layers at the beginning and the ending. 

6) \textbf{Graph Element Network (GEN)} \citep{alet2019graph}: Neural operator with graph convolution backbone.

7) \textbf{Spectral Neural Operator (SNO)} \citep{fanaskov2022spectral}: linear combination of a fixed basis (Eq.~\ref{eq: spectral method}), with neuralized coefficients. It is a degenerated version of NASM and will be compared in the ablation study part.

\begin{figure*}[t]
    \centering
    \begin{subfigure}[b]{0.33\textwidth}
        \centering
        \includegraphics[width=\textwidth]{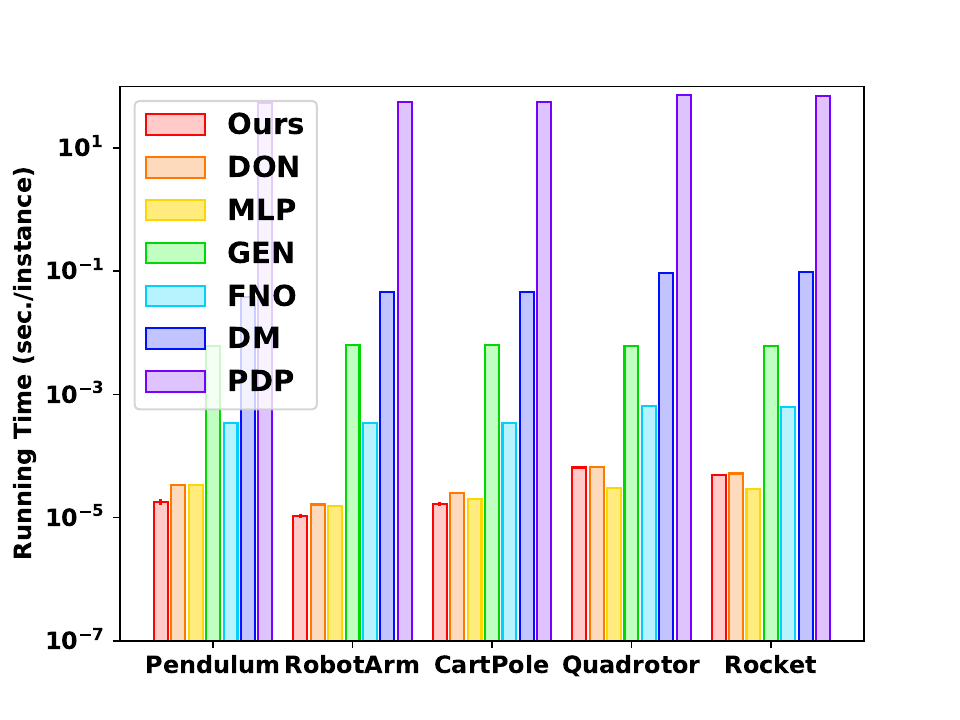}
        \caption{ Time (sec./instance)}
        \label{subfig:time}
    \end{subfigure}\hfil
    \begin{subfigure}[b]{0.33\textwidth}
         \centering
         \includegraphics[width=\textwidth]{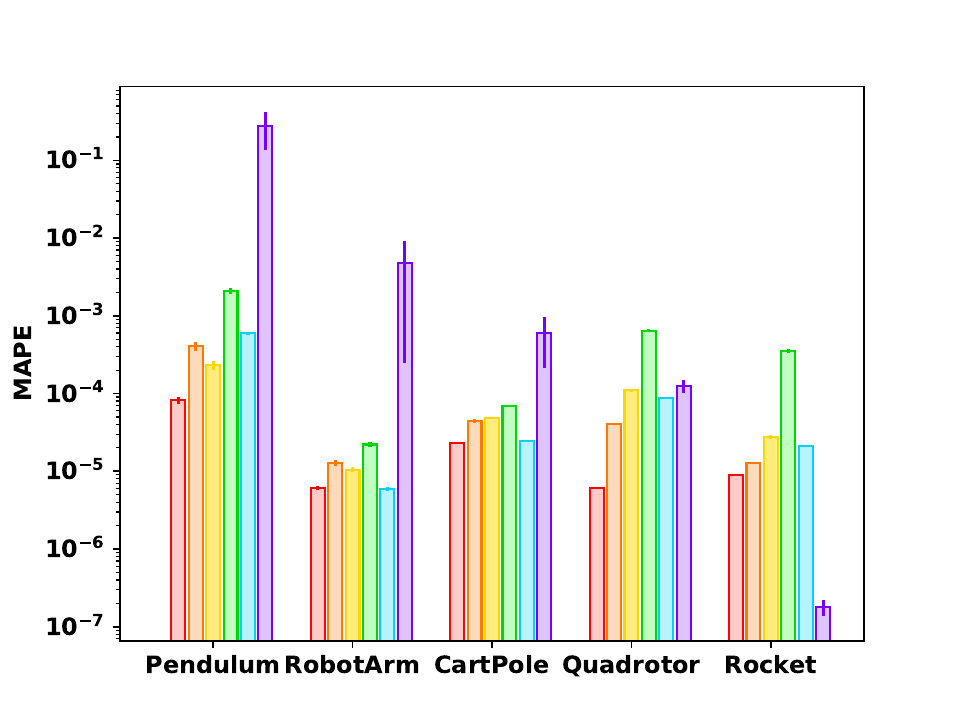}
         \caption{ID MAPE}
         \label{subfig:inErr}
     \end{subfigure}\hfil
    \begin{subfigure}[b]{0.33\textwidth}
         \centering
         \includegraphics[width=\textwidth]{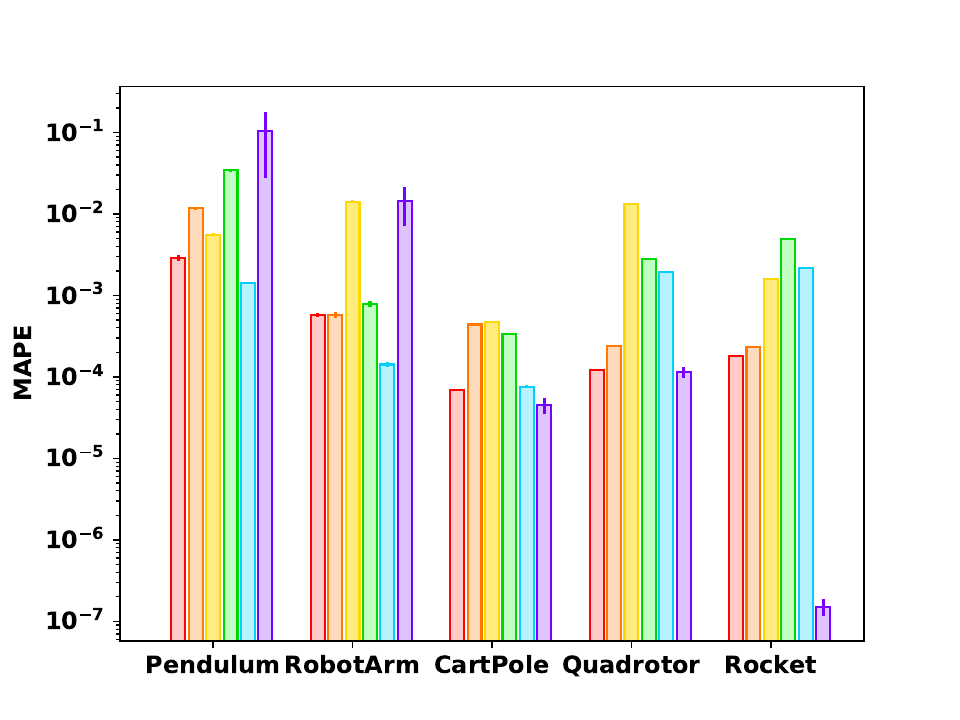}
         \caption{OOD MAPE}
         \label{subfig:outErr}
     \end{subfigure}
     
    \caption{Inference time and mean absolute percentage error (MAPE) on in-distribution (ID) and OOD benchmarks.  NASM (red bars) achieves higher or comparable accuracy, with the fastest or second fastest speed.}
    \label{fig:summary}
\end{figure*}
\subsubsection{Results and Discussion}
We present the numerical results on the five systems to evaluate the efficiency and accuracy of NASM and other architectures for instance-solution operator framework. The metrics of interest are 1) the running time of solving problems; 2) the quality of solution, measured by mean absolute percentage error (MAPE) between the true optimal cost and the predicted cost, which is defined as the mean of $|(J_{opt} - J_{sol})/J_{opt}|$, where $J_{opt}$ is the optimal cost generated by DM (regarded as ground truth), and $J_{sol}$ is the cost of the solution produced by the model. The MAPE is calculated on ID/OOD benchmarks respectively, and the running time is averaged for 2,000 random problems. The results of ODE-constrained OCP are visualized in Fig.~\ref{fig:summary}.

First, the comparison of the wall-clock running time is shown in Fig.~\ref{subfig:time} as well as in the third column of Tab.~\ref{tab:Quadrotor}, which shows that the neural operator solvers are much faster than the classic solvers, although they both tested on CPU for fairness. For example, \name achieves over 6000 times speedup against the DM solver. The acceleration can be reasoned in two aspects: 1) the neural operator solvers produce the output by a single forward propagation, while the classic methods need to iterate between forward and backward pass multiple times; 2) the neural solver calculation is highly paralleled. Among the neural operator solvers, the running time of NASM, DON, and MLP are similar. The FNO follows a similar diagram as MLP, but 10+ times slower than MLP, since it involves computationally intensive Fourier transformations. The GEN is 100+ times slower than MLP, probably because of the intrinsic complexity of graph construction and graph convolution.

\begin{table*}[tb!]
\centering
\caption{Results of Quadrotor environment.}
\label{tab:Quadrotor}
\begin{tabular}{@{}llccc@{}}
\toprule
Formulation & Model    & Time(sec./instance) & ID MAPE  & OOD MAPE \\ \midrule
Direct Method& DM & $9.23 \times 10^{-2}$ & $\diagdown$  & $\diagdown$ \\
Indirect Method& PDP & $7.25 \times 10^{1}$ & $1.24 \times 10^{-4}$  & $1.16 \times 10^{-4}$ \\\midrule
\multirow{5}{*}{\shortstack[l]{Instance-Solution\\
Control Operator}}& NASM & $6.50 \times 10^{-5}$ & $\mathbf{6.17 \times 10^{-6}}$  & $\mathbf{1.21 \times 10^{-4}}$ \\
& DON & $6.54 \times 10^{-5}$ & $4.09 \times 10^{-5}$  & $2.40 \times 10^{-4}$ \\
& MLP & $\mathbf{3.04 \times 10^{-5}}$ & $1.10 \times 10^{-4}$  & $1.33 \times 10^{-2}$ \\
& GEN & $6.15 \times 10^{-3}$ & $6.40 \times 10^{-4}$  & $2.77 \times 10^{-3}$ \\
& FNO & $6.38 \times 10^{-4}$ & $8.73 \times 10^{-5}$  & $1.92 \times 10^{-3}$ \\
\bottomrule
\end{tabular}
\end{table*} 
The accuracy on in- and out-of-distribution benchmark sets is compared in Fig.~\ref{subfig:inErr}-\ref{subfig:outErr}. Compared with other neural models, \name achieves better or comparable accuracy in general. In addition, \name outperforms classical PDP on more than half of the benchmarks. As a concrete example, we investigate the performance of the Quadrotor environment. From Tab.~\ref{tab:Quadrotor}, one can observe that MAPE of \name on the ID and OOD is the best among all neural models. The OOD performance is close to that of the classical method PDP, which is insensitive to distribution shift.  We conjecture that both ID/OOD generalization ability of \name results from its architecture, where coefficients and basis functions are explicitly disentangled. Such a structure may inherit the inductive bias from numerical basis expansion methods (e.g. \cite{kafash2014numerical}), thus being more robust to distribution shifts.  

For NASM in all synthetic environment experiments, we choose MLP as the backbone of Coefficient Net, Fourier adaptive basis, non-static coefficients, and summation aggregation. To justify these architecture choices, we perform an ablation study by changing or removing each of the modules, as displayed in Tab.~\ref{tab:archi-just}. We use the abbreviation w.o. for 'without', and w. for 'with'. The first row denotes the unchanged model, and the second and third rows are fixing basis functions, and using static (time-independent) Coefficient Net. These two modifications slightly lower the performance in both ID and OOD. The 4th row is Fourier SNO, i.e. applying the above two modifications together on NASM, which dramatically increases the error. The reason is that SNO uses a fixed Fourier basis, and thus requires the optimal control to be globally periodic for accurate interpolation, while Quadrotor control is non-periodic. If the basis of SNO changes to non-periodic, such as Chebyshev basis as in the 5th row, then the performance is covered. The result shows that the accuracy of SNO depends heavily on the choice of pre-defined basis function. In comparison, NASM is less sensitive to the choice of basis due to its adaptive nature, as the 6th row shows that changing the basis from Fourier to Chebyshev hardly changes the NASM's performance. The 7th and 8th rows demonstrate that both replacing summation with a neural network aggregation and switching to a Convolutional network backbone do not improve the performance on Quadrotor.

\begin{table}[t!]
\centering
\caption{Architecture justification of NASM on Quadrotor.}
\label{tab:archi-just}
\begin{tabular}{@{}clccc@{}}
\toprule
& Modification    & ID MAPE  & OOD MAPE \\ \midrule
1& NASM  & $\mathbf{6.17 \times 10^{-6}}$  & $\mathbf{1.21 \times 10^{-4}}$ \\
2& w.o. Adapt Params & $7.86 \times 10^{-6}$  & $1.97 \times 10^{-4}$ \\
3& w.o. Non-static Coef  & $9.01 \times 10^{-6}$  & $5.83 \times 10^{-4}$ \\ \midrule
4& SNO (Fourier)  & $9.59 \times 10^{-2}$  & $8.39 \times 10^{-2}$ \\
5& SNO (Chebyshev)  & $6.64 \times 10^{-6}$  & $3.82 \times 10^{-4}$ \\ 
6& w.   Chebyshev  & $6.36 \times 10^{-6}$  & $2.86 \times 10^{-4}$ \\ \midrule
7& w.   Neural Aggr  & $7.13 \times 10^{-6}$  & $1.29 \times 10^{-4}$ \\
8& w.   Conv  & $1.16 \times 10^{-5}$  & $4.95 \times 10^{-3}$ \\
\bottomrule
\end{tabular}
\end{table} 


\subsubsection{Extended Experiment Result on Quadrotor}
\textbf{1) More Variables.} In the previous experiments, only the cost function (target state) is changeable. Here, we add more variables to the network input, such that the cost function, dynamics (physical parameters e.g. wing length), and initial condition are all changeable. The in/out- distributions for dynamics and initial conditions are uniform distributions, designed similarly to that of the target state. The result is listed in Tab.~\ref{tab:Quadrotor-dit}.
\begin{table}[tb!]
\centering
\caption{Quadrotor with changeable cost, dynamics and initial conditions.}
\label{tab:Quadrotor-dit}
\begin{tabular}{@{}lccc@{}}
\toprule
Model    & Time(sec./instance) & ID MAPE  & OOD MAPE \\ \midrule
DM & $1.59 \times 10^{-1}$ & $\diagdown$  & $\diagdown$ \\
PDP & $7.25 \times 10^{1}$ & $1.13 \times 10^{-4}$  & $1.15 \times 10^{-4}$ \\ \midrule
\name & $4.40 \times 10^{-5}$ & $\mathbf{5.92 \times 10^{-5}}$  & $\mathbf{1.57 \times 10^{-4}}$ \\
DON & $\mathbf{3.39 \times 10^{-5}}$ & $1.02 \times 10^{-4}$  & $3.19 \times 10^{-4}$ \\
MLP & $8.45 \times 10^{-5}$ & $5.94 \times 10^{-4}$  & $2.36 \times 10^{-3}$ \\
GEN & $1.10 \times 10^{-2}$ & $6.06 \times 10^{-4}$  & $2.32 \times 10^{-3}$ \\
FNO & $1.95 \times 10^{-3}$ & $6.22 \times 10^{-4}$  & $2.32 \times 10^{-3}$ \\
\bottomrule
\end{tabular}
\end{table}

\textbf{2) Extreme OOD shifts and Transfer Learning.}
In addition to the OOD shifts in the previous experiments, we design some more extremely shifted distributions:
$\vb*{\epsilon}_{out}^1  \sim \mathcal{U}(-1.0, -0.8), \vb*{\epsilon}_{out}^2  \sim \mathcal{U}(-1.3, -1.1), \vb*{\epsilon}_{out}^3  \sim \mathcal{U}(-1.6, -1.4)$

\begin{table}[tb!]
\centering
\caption{MAPE of Quadrotor over OOD shift, without finetune.}
\label{tab:Quadrotor without finetune}
 
\begin{tabular}{@{}lccc@{}}
\toprule
Model &  $\vb*{\epsilon}_{out}^1$& $\vb*{\epsilon}_{out}^2$& $\vb*{\epsilon}_{out}^3$\\ \midrule
\name &  $\mathbf{1.29 \times 10^{-3}}$ & $6.90 \times 10^{-3}$ & $1.50 \times 10^{-2}$\\
DON & $2.32 \times 10^{-3}$ & $7.58 \times 10^{-3}$ & $1.61 \times 10^{-2}$\\
MLP &  $3.15 \times 10^{-3}$ & $\mathbf{6.10 \times 10^{-3}}$ & $\mathbf{9.53 \times 10^{-3}}$\\
GEN &  $1.14 \times 10^{-2}$ & $3.94 \times 10^{-2}$ & $2.18 \times 10^{-1}$\\
FNO &  $1.35 \times 10^{-2}$ & $5.04 \times 10^{-2}$ & $1.03 \times 10^{-1}$\\
\bottomrule
\end{tabular}
\end{table} 

Tab.~\ref{tab:Quadrotor without finetune} shows the OOD MAPE on those distributions. From the table, one can observe that the performance of neural models decreases with more distribution shifts. When the distribution shift increases to $\vb*{\epsilon}_{out}^3$, the results of neural models are almost unacceptable.


In practice, if large shift OOD reusability is required for some applications, we may assume a few training samples from OOD are available (i.e. few-shot). NCOs can be fine-tuned on the training samples from OOD to achieve better performance. This idea has been proposed for DeepONet transfer learning in \citep{goswami2022deep}.

We follow the fine-tuning scheme proposed in \citep{goswami2022deep}. We set the size of the OOD fine-tuning dataset and \#epochs to be 20\% of that of the ID training dataset. The learning rate is fixed at 0.001. The embedding network, basis function net, and the first two convolution blocks are all frozen during fine-tuning since we assume they keep the distribution invariant information.

\begin{table}[tb!]
\centering
\caption{MAPE of Quadrotor on various OOD shift, with finetune.}
\label{tab:Quadrotor with finetune}
 
\begin{tabular}{@{}lcccc@{}}
\toprule
Model & $\vb*{\epsilon}_{out}^1$& $\vb*{\epsilon}_{out}^2$& $\vb*{\epsilon}_{out}^3$\\ \midrule
\name  & $\mathbf{1.02 \times 10^{-5}}$ & $\mathbf{2.29 \times 10^{-5}}$ & $3.07 \times 10^{-5}$\\
DON  & $1.95 \times 10^{-5}$ & $2.57 \times 10^{-5}$ & $\mathbf{3.02 \times 10^{-5}}$\\
MLP  & $3.10 \times 10^{-5}$ & $1.01 \times 10^{-4}$ & $1.88 \times 10^{-4}$\\
GEN  & $2.36 \times 10^{-4}$ & $4.07 \times 10^{-4}$ & $8.06 \times 10^{-4}$\\
FNO  & $4.34 \times 10^{-5}$ & $1.16 \times 10^{-4}$ & $2.11 \times 10^{-4}$\\
\bottomrule
\end{tabular}
\end{table} 

Tab.~\ref{tab:Quadrotor with finetune} reveals that fine-tuning greatly improves neural model performance on extreme distribution shifts. Therefore, the availability of fine-tuning on extra data benefits the reusability of neural operators on OCP.

\subsection{Real-world Dataset of Planar Pushing}
\label{sec:pushing}
\begin{figure}[t]
\centering
\includegraphics[width = 0.7\linewidth]{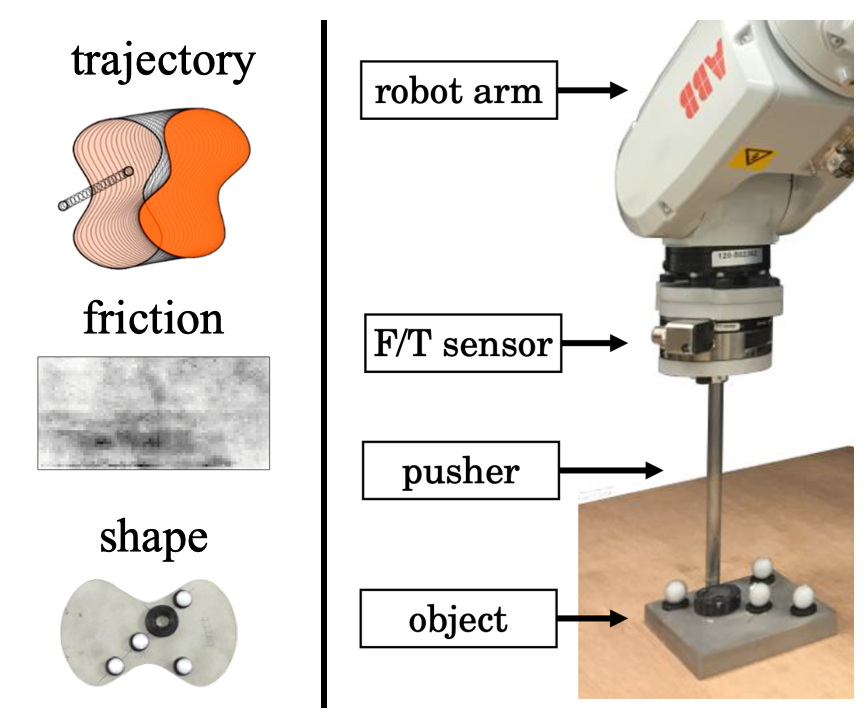}
\captionof{figure}{Pushing environment\cite{yu2016more}.}
\label{fig:push-data}
\end{figure}

We present how to learn OC of a robot arm for pushing objects of varying shapes on various planar surfaces. We use the \emph{Pushing} dataset \cite{yu2016more}, a challenging dataset consisting of noisy, real-world data produced by an ABB IRB 120 industrial robotic arm (Fig.~\ref{fig:push-data}, right part). 

The robot arm is controlled to push objects starting from various contact positions and 9 angles (initial state), along different trajectories (target state functions), with 11 object shapes and 4 surface materials (dynamics). The control function is represented by the force exerted on to object. Left of Fig.~\ref{fig:push-data} gives an overview of input variables.  

We apply \name to learn a mapping from a pushing OCP instance (represented by variables above) to the optimal control function. The input now is no longer the parameters of cost functional $f$ only, but parameters and representations of $(f, \vb*{d}, \vb*{x}_{init})$. The encoder is realized by different techniques for different inputs, such as Savitzky–Golay smoothing \citep{savitzky1964smoothing} and down-sampling for target trajectories, mean and standard value extraction for friction map, and CNN for shape images. 

We extract training data from ID, validation, and test data from both ID/OOD. The ID/OOD is distinguished by different initial contact positions of the arm and object. The accuracy metric MAPE is now defined as $\norm{\hat{u} - u^*}/\norm{u^*}$. We compare \name performance only with neural baselines since the explicit expression of pushing OCP is unavailable, thus classical methods DM and PDP are not applicable. All neural models share the same encoder structure, while the parameters of the encoder are trained end-to-end individually for each model. The results are displayed in Tab.~\ref{tab:Pushing}, from which one can observe that \name outperforms all baselines in the ID MAPE and that \name achieves the second lowest OOD MAPE, with a slight disadvantage against DON.  

The design of NASM in the pushing environment is an adaptive Chebyshev basis, non-static Coefficient Net with Convolution backbone, and neuralized aggregation. The justification of the design is shown in Tab.~\ref{tab:Pushing}. Its performance is insensitive to the choice of basis, as the second row reveals. Both the convolution backbone and the neuralized aggregation contribute to the high accuracy. A possible reason is that the control operator $\mathcal{G}$ in the pushing dataset is highly non-smooth and noisy, thus a more complicated architecture with higher capability and flexibility is preferred.

\begin{table}
\centering
\captionof{table}{Results of Pushing environment.}
\label{tab:Pushing}
\begin{tabular}[b]{@{}lcc@{}}
\toprule
Model    & ID MAPE  & OOD MAPE \\ \midrule
NASM & \textbf{0.108} ($\pm$ 0.006) & 0.137 ($\pm$ 0.007) \\
NASM(Fourier) & 0.113 ($\pm$ 0.006) & 0.139 ($\pm$ 0.009) \\
NASM(w.o. Conv) & 0.132 ($\pm$ 0.008) & 0.151 ($\pm$ 0.009) \\
NASM(w.o. NAggr) & 0.125 ($\pm$ 0.007) & 0.172 ($\pm$ 0.009) \\
\midrule
DON & 0.117 ($\pm$ 0.006) & \textbf{0.134} ($\pm$ 0.008) \\
MLP  & 0.128 ($\pm$ 0.006) & 0.149 ($\pm$ 0.008) \\
GEN  & 0.209 ($\pm$ 0.014) & 0.199 ($\pm$ 0.015) \\
FNO  & 0.131 ($\pm$ 0.008) & 0.172 ($\pm$ 0.009) \\
\bottomrule
\end{tabular}
\end{table}

\section{Conclusion and Outlook}
We have proposed an instance-solution operator perspective of OCPs, where the operator directly maps cost functionals to OC functions. We then present a neural operator \name, with a theoretic guarantee on its approximation capability. Experiments show outstanding generalization ability and efficiency, in both ID and OOD settings. We envision the proposed model will be beneficial in solving numerous high-dimensional problems in the learning and control fields. 

We currently do not specify the forms of problem instances (e.g. ODE/PDE-constrained OCP) or investigate sophisticated models for specific problems, which calls for careful designs and exploitation of the problem structures. 

\begingroup
\fontsize{9}{10}\selectfont
\bibliography{aaai25}
\endgroup


\onecolumn
\appendix

\section{Algorithms} 
\label{appx: algorithms}
The data generation algorithm is listed in Alg. \ref{alg:data gen}.
\begin{algorithm*}[tb!]
\caption{Data generation}
\label{alg:data gen}
\textbf{Input}: Distribution of target state $\vb{x}_{goal}$ \\
\textbf{Output}: Dataset
\begin{algorithmic}[1]
\STATE $N \leftarrow$ Number of samples per trajectory; 
\STATE Dataset $\leftarrow$ empty set;
\WHILE{not Dataset is full}
    \STATE sample $\vb*{x}_{goal}$ from $\vb{x}_{goal}$\;
    \STATE construct cost functional $f$  with $\vb*{x}_{goal}$\;
    \STATE construct OCP (Eq. \ref{eq:oc}) with cost functional $f$\;
    \STATE $\vb*{u}^* \leftarrow$ OC\_Solver(OCP); \COMMENT{Any solver is applicable. We choose DM.}
    \STATE sample $\{t_j\}_{1\le j \le N}$ from time horizon $T$ of OCP\;
    \STATE add triplets $\{(f, t_j, \vb*{u}^*(t_j)) \}_{1\le j \le N}$ to Dataset\;
\ENDWHILE
\STATE \textbf{return} Dataset
\end{algorithmic}

\end{algorithm*}

\section{Related Works}
\label{appx: related works}
\subsection{OCP Solvers}
Traditional numerical solvers are well developed over the decades, which are learning-free and often involve tedious optimization iterations to find an optimal solution. 


\textbf{Direct methods}~\citep{Böhme2017direct} reformulate OCP as finite-dimensional nonlinear programming (NLP) \citep{bazaraa2013nonlinear}, and solve the problem by NLP algorithms, e.g. sequential quadratic programming \citep{boggs1995sequential} and interior-point method \citep{mehrotra1992implementation}. The reformulation essentially constructs surrogate models, where the state and control function (infinite dimension) is replaced by polynomial or piece-wise constant functions. The dynamics constraint is discretized into equality constraints. The direct methods optimize the surrogate models, thus the solution is not guaranteed to be optimal for the origin problem. Likewise, typical direct neural solvers ~\citep{chen2018optimal, wang2021fast, hwang2022solving}, termed as Two-Phase models, consist of two phases: 1) approximating the dynamics by a neural network (surrogate model); 2) solving the NLP via gradient descent, by differentiating through the network. The advantage of Two-Phase models against traditional direct methods is computational efficiency, especially in high-dimensional cases. However, the two-phase method is sensitive to distribution shift (see Fig.~\ref{fig:PendulumTP}).

\textbf{Indirect methods}~\citep{Böhme2017indirect} are based on Pontryagin’s Maximum Principle (PMP)~\citep{pontryagin1987mathematical}. By PMP, indirect methods convert OCP (Eq.~\ref{eq:oc}) into a boundary-value problem (BVP)~\citep{lasota1968discrete}, which is then solved by numerical methods such as shooting method \citep{bock1984multiple}, collocation method \citep{xiu2005high}, adjoint-based gradient descend \citep{effati2013optimal,jin2020pontryagin}. These numerical methods are sensitive to the initial guesses of the solution. Some indirect methods-based neural solvers approximate the finite-dimensional mapping from state $\vb*{x}^\ast(t) \in \mathbb{R}^{d_x}$ to control $\vb*{u}^\ast(t)\in \mathbb{R}^{d_u}$ \citep{cheng2020real}, or to co-state $\vb*{\lambda}^\ast(t) \in \mathbb{R}^{d_x}$ \citep{xie2018artificial}. The full trajectory of the control function is obtained by repeatedly applying the mapping and getting feedback from the system, and such sequential nature is the efficiency bottleneck. Another work \citep{d2021pontryagin} proposes to solve the BVP via a PINN, thus its trained network works only for one specific OCP instance.

\textbf{Dynamic programming (DP)} is an alternative, based on Bellman's principle of optimality \citep{bellman1960dynamic}. It offers a rule to divide a high-dimensional optimization problem with a long time horizon into smaller, easier-to-solve auxiliary optimization problems. Typical methods are Hamilton-Jacobi-Bellman (HJB) equation \citep{al2008discrete}, differential dynamical programming (DDP) \citep{tassa2014control}, which assumes quadratic dynamics and value function, and iterative linear quadratic regulator (iLQR)~\citep{li2004iterative}, which assumes linear dynamics and quadratic value function. Similar to dynamic programming, model predictive control (MPC) synthesizes the approximate control function via the repeated solution of finite overlapping horizons~\citep{hewing2020learning}. The main drawback of DP is the \emph{curse of dimensionality} on the number and complexity of the auxiliary problem. MPC alleviates this problem at the expense of optimality. Yet fast implementation of MPC is still under exploration and remains open~\citep{nubert2020safe}.

\subsection{Differential Equation Neural solvers}
A variety of networks have been developed to solve DE, including Physics-informed neural networks (PINNs) \citep{raissi2019physics}, neural operators \citep{lu2021learning}, hybrid models \citep{mathiesen2022hyperverlet}, and frequency domain models \citep{li2020fourier}. We will briefly introduce the first two models for their close relevance to our work. 

\textbf{PINNs} parameterize the DE solution as a neural network and learn the parameters by minimizing the residual loss and boundary condition loss \citep{yu2018deep, raissi2019physics}. PINNs are similar to those numerical methods e.g. the finite element method in that they replace the linear span of a finite set of local basis functions with neural networks. PINNs usually have simple architectures (e.g. MLP), although they have produced remarkable results across a wide range of areas in computational science and engineering \citep{raissi2020hidden, zhu2019physics}. However, these models are limited to learning the solution of one specific DE instance, but not the operator. In other words, if the coefficients of the DE instance slightly change, then a new neural network needs to be trained accordingly, which is time-consuming. Another major drawback of PINNs, as pointed out by \citep{wang2021understanding}, is that the magnitude of two loss terms (i.e.residual loss and boundary condition loss) is inherently imbalanced, leading to heavily biased predictions even for simple linear equations.

\textbf{Neural operators} regards DE as an operator that maps the input to the solution. Learning operators using neural networks was introduced in the seminal work \citep{chen1995universal}. It proposes the universal approximation theorem for operator learning, i.e. a network with a single hidden layer can approximate any nonlinear continuous operator. \citep{lu2021learning} follows this theorem by designing a deep architecture named \emph{DeepONet}, which consists of two networks: a branch net for input functions and a trunk net for the querying locations in the output space. The DeepONet output is obtained by computing the inner product of the outputs of branch and trunk networks. The DeepONet can be regarded as a special case of our \name, since DeepONet is equivalent to a \name without the convolution blocks and the projecting network.  And our analysis of optimal control error bound is partly inspired by~\citep{lanthaler2022error}, providing error estimation of DeepONet. 

In addition, there exist many neural operators with other architectures. For example, another type of neural operator is to parameterize the operator as a convolutional neural network (CNN) between the finite-dimensional data meshes \citep{zhu2018bayesian, khoo2021solving}. The major weakness of these models is that it is impossible to query solutions at off-grid points. Moreover, graph neural networks (GNNs)~\citep{kipf2016semi} are also applied in operator learning \citep{alet2019graph, anandkumar2020neural}. The key idea is to construct the spacial mesh as a graph, where the nodes are discretized spatial locations, and the edges link neighboring locations. Compared with CNN-based models, the graph operator model is less sensitive to resolution and is capable of inferencing at off-grid points by adding new nodes to the graph. However, its computational cost is still high, growing quadratically with the number of nodes. Another category of neural operators is Fourier transform based \citep{anandkumar2020neural,li2020fourier}.  The models learn parametric linear functions in the frequency domain, along with nonlinear functions in the time domain. The conversion between those two domains is realized by discrete Fourier transformation. The approximator architecture of \name is inspired by the Fourier neural operators, and we replace the matrix product in the frequency domain with the convolution in the original domain.

\section{Approximation Error Theorems}
\label{appx: appx err thoerem}
We give the estimation for the approximation error of the proposed model. The theoretic result guarantees that  
there exists a neural network instance of \name architecture approximating the operator $\mathcal{G}$ to arbitrary error tolerance. Furthermore, the size and depth of such a network are upper-bounded. The error bound will be compared with the DON error bound derived in \citep{lanthaler2022error}. The technical line of our analysis is inspired by \citep{lanthaler2022error} which provides error estimation for DON's three components: encoder, approximation, and reconstructor.

\subsection{Definition and Decomposition of Errors}
For fair and systematic comparison, we also decompose the architecture of NASM into three components. The encoder aims to convert the infinite-dimensional input to a finite vector, as introduced in the main text. The approximator of NASM is the coefficient network that produces the (approximated) coefficients. The reconstructor is the basis functions and the aggregation, that reconstructs the optimal control via the coefficients. In this way, the approximation error of \name can also be decomposed into three parts: 1) encoder error, 2) approximator error, and 3) reconstructor error. 

For ease of analysis, the error of the encoder is temporally not considered and is assumed to be zero. The reason is that the design of the encoder is more related to the problem setting instead of the neural operator itself. In our experiment, all neural operators use the same encoder, thus the error incurred by the encoder should be almost the same. Therefore, assuming the zero encoder error is a safe simplification. In our assumption, for a given encoder $\mathcal{E}$, there exists an inverse mapping $\mathcal{E}^{-1}$, such that $\mathcal{E}^{-1} \circ \mathcal{E} = \operatorname{Id}$.  The non-zero encoder error is analyzed in Appx \ref{appx:extension}.

For a reconstructor $\mathcal{R}$, its error $\widehat{\mathscr{E}}_{\mathcal{R}}$ is estimated by the mismatch between $\mathcal{R}$ and its approximate inverse mapping, projector $\mathcal{P}$, weighted by push-forward measure $\mathcal{G}_{\#} \mu (u) := \mu (\mathcal{G}^{-1}(u))$. 
\begin{equation}
\begin{aligned}
\widehat{\mathscr{E}}_{\mathcal{R}} &:=& \left(\int_{U} \norm{\mathcal{R} \circ \mathcal{P}(u)-u}_U \dd \left(\mathcal{G}_{\#} \mu\right)(u)\right)^{\frac{1}{2}} \\
\mathcal{P} &:=& \operatornamewithlimits{argmin}_{\mathcal{P}} \widehat{\mathscr{E}}_{\mathcal{R}}, \quad \text{s.t.} \quad \mathcal{P} \circ \mathcal{R}=\mathrm{Id}.
\end{aligned}
\end{equation}
Intuitively, such reconstructor error quantifies the information loss induced by $\mathcal{R}$. An ideal $\mathcal{R}$ without any information loss should be invertible, i.e. its optimal inverse $\mathcal{P}$ is exactly $\mathcal{R}^{-1}$, thus we have $\widehat{\mathscr{E}}_{\mathcal{R}}=0$.

Given the encoder and reconstructor, and denote the encoder output is $\vb*{e} \in \mathbb{R}^m$, the error $\widehat{\mathscr{E}}_{\mathcal{A}}$ induced by approximator $\mathcal{A}$ is defined as the distance between the approximator output and the optimal coefficient vector, weighted on push-forward measure $\mathcal{E}_{\#} \mu(\vb*{e}) := \mu(\mathcal{E}^{-1}(\vb*{e}) ) $:
\begin{equation}
\widehat{\mathscr{E}}_{\mathcal{A}}:=\left(\int_{\mathbb{R}^{m}}\|\mathcal{A}(\vb*{e})-\mathcal{P} \circ \mathcal{G} \circ \mathcal{E}^{-1}(\vb*{e})\|_{\ell^{2}\left(\mathbb{R}^{p}\right)}^{2} \dd \left(\mathcal{E}_{\#} \mu\right)(\vb*{e})\right)^{\frac{1}{2}}.
\label{eq:approximator-error}
\end{equation}

With the definitions above, we can estimate the approximation error of our \name by the error of each of its components, as stated in the following theorem (see detailed proof in Appendix \ref{pf:decomp approx error}):
\begin{theorem}[\textbf{Decomposition of \name Approximation Error}] Suppose the cost functional $f_{i}$ is Lipschitz continuous, with Lipschitz constant $\operatorname{Lip}(f_{i})$. Define constant 
$
C = \sup_{i \in I} \frac{\operatorname{Lip}(f_{i})}{f_{i} \circ \mathcal{G}(i)}.
$ The approximation error $\widehat{\mathscr{E}}$ (Eq. \ref{eq:approx err}) of a \name $\mathcal{N}=\mathcal{R} \circ \mathcal{A} \circ \mathcal{E}$ is upper-bounded by
\begin{equation}
    \widehat{\mathscr{E}} \leq C \qty(\operatorname{Lip}(\mathcal{R}) \widehat{\mathscr{E}}_{\mathcal{A}}+\widehat{\mathscr{E}}_{\mathcal{R}} ).
\end{equation}
\label{th:decomp approx error}
\end{theorem}

\subsection{Estimation of Decomposed Errors}
The reconstructor error $\widehat{\mathscr{E}}_{\mathcal{R}}$ of NASM reduces to the error of interpolation by basis functions, which is thoroughly studied in approximation theory \citep{davis1975interpolation, powell1981approximation}. For the Fourier basis, we restate the existing result to establish the following theorem:
\begin{theorem} [\textbf{Fourier Reconstructor Error}] 
If $\mathcal{G}$ defines a Lipschitz mapping $\mathcal{G}: I \rightarrow H^{s}(T)$ ,for some $s>0, M>0$, with
$
\int_{I}\|\mathcal{G}(i)\|_{H^{s}}^{2} \dd \mu(i) \leq M<\infty,
$
then there exists a constant $C=C(s, M)>0$, such that for any $p \in \mathbb{N}$, the Fourier basis $\vb*{b}: T \rightarrow \mathbb{R}$ with the summation aggregation and the associated reconstruction $\mathcal{R}: \mathbb{R}^{p} \rightarrow U$ satisfies:
\begin{eqnarray}
 \widehat{\mathscr{E}}_{\mathcal{R}} \leq C p^{-s}.
\label{eq:approximation-error}
\end{eqnarray}

Furthermore, the reconstruction $\mathcal{R}$ satisfy $\operatorname{Lip}(\mathcal{R}) \leq 1$.
\end{theorem}

 $H^{s}$ is a Sobolev space with $s$ degrees of regularity and $L^2$ norm. The proof (omitted here) is based on an observation that a smooth function (i.e. with large $s$) changes slowly with time, having small (exponentially decaying) coefficients in the high frequency. 

Next, the error bound of the approximator will be presented. Notice that the approximator learns to map between two vector spaces using MLP, whose error estimation is well-studied in deep learning theory. One of the existing works \citep{guhring2020error} derives the estimation based on the Sobolev regularity of the mapping. We extend their result to both MLP and Convolution (Appendix \ref{appx: error mlp cnn}). And the approximator error can be directly obtained from those results. We present the MLP-backbone approximator error as follows, and the Convolution-backbone error can be similarly derived.
\begin{theorem}[\textbf{Approximator Error (MLP backbone)}] 
\label{th:appx error}
 Given operator $\mathcal{G}: I \rightarrow U$, encoder $I \rightarrow \mathbb{R}^m$, and reconstructor $\mathcal{R}: \mathbb{R}^p \rightarrow U$, let $\mathcal{P}$ denote the corresponding projector. If for some $s \in \mathbb{N}_{\ge 2}, M>0$, the following bound is satisfied:
 $
 \norm{\mathcal{P} \circ \mathcal{G} \circ \mathcal{E}^{-1}}_{H^{s}(\mathcal{E}_{\#} \mu)} \le M < \infty,
 $
then there exists a constant $C=C(m, s, M)>0$, such that for any $\varepsilon \in (0, \frac{1}{2})$, there exists an approximator $\mathcal{A}: \mathbb{R}^m \rightarrow \mathbb{R}^p$ implemented by MLP such that:
\begin{eqnarray}
\operatorname{size}(\mathcal{A}) \leq  C p^2\varepsilon^{-m /s}  \log \left(\varepsilon^{-1}\right), \qquad
\operatorname{depth}(\mathcal{A}) \leq C  \log \left(\varepsilon^{-1}\right), \qquad
\widehat{\mathscr{E}}_{\mathcal{A}} \leq \sqrt{p} \varepsilon.
\end{eqnarray}
\end{theorem}

Note that $\operatorname{size}(\cdot )$ is defined as the number of trainable parameters of a neural network, and $\operatorname{depth}(\cdot )$ denotes the number of hidden layers.

In summary, we have proved that the approximation error is bounded by the sum of the reconstructor and approximator errors. Those errors can hold under arbitrary small tolerance, with bounded size and depth. The theorems hold as long as some integrative and continuous conditions are satisfied \citep{dontchev2006well}, which is trivial in many real-world continuous OCPs.

\subsection{Comparasion with DeepONet Error}
\label{appx: comp with don}

For fairness, we assume the DON uses the same encoder and network backbone as the NASM. Then the approximator error and the encoder error of both operators should be the same. We only need to compare the reconstructor error.

The error of DeepONet Reconstructor $\mathcal{R}_{don}$ is analyzed in a previous work \citep{lanthaler2022error}, by assuming the DeepONet trunk-net learns an approximated Fourier basis. We cite the result to establish the following theorem:
\begin{theorem} [\textbf{DeepONet Reconstructor Error}~\citep{lanthaler2022error}] 
If $\mathcal{G}$ defines a Lipschitz mapping $\mathcal{G}: I \rightarrow H^{s}(T)$ ,for some $s>0, M>0$, with
$\norm{\mathcal{G}}_{H^{s}}^{2}  \leq M<\infty$, 
then there exists a constant $C=C(s, M)>0$, such that for any $p \in \mathbb{N}$, there exists a trunk net (i.e. the basis function net) $\boldsymbol{\tau}: T \rightarrow \mathbb{R}^{p}$  and the associated reconstruction $\mathcal{R}_{don}: \mathbb{R}^{p} \rightarrow U$ satisfies:
\begin{eqnarray}
\operatorname{size}(\mathcal{R}_{don}) &\leq &  C p\left(1+\log (p)^{2}\right), \\
\operatorname{depth}(\mathcal{R}_{don}) &\leq & C\left(1+\log (p)^{2}\right), \\
\widehat{\mathscr{E}}_{\mathcal{R}_{don}} &\leq & C p^{-s}.
\end{eqnarray}

Furthermore, the reconstruction $\mathcal{R}_{don}$ and the optimal projection $\mathcal{P}_{don}$ satisfy $\operatorname{Lip}(\mathcal{R}_{don}), \operatorname{Lip}(\mathcal{P}_{don}) \leq 2$.
\end{theorem}

One can observe that the DON reconstructor error has the same form as that of NASM. However, the DON reconstructor requires a neural network (trunk-net) to implement basis functions, while NASM uses a pre-defined Fourier basis. In the degenerated form of the NASM, i.e. without adaptive parameters and neural aggregations, the NASM reconstructor has no learnable parameters. In this sense, NASM can achieve the same error bound with fewer parameters, compared with DON.  

\section{Mathematical Preliminaries and Proofs}
\label{appx:proofs}

\subsection{Preliminaries on Sobolev Space}
For $d \geq 1, \Omega$ an open subset of $\mathbb{R}^d, p \in[1 ;+\infty]$ and $s \in \mathbb{N}$, the Sobolev space $W^{s, p}\left(\mathbb{R}^d\right)$ is defined by
$$
W^{s, p}(\Omega)=\left\{f \in L^p(\Omega): \forall|\alpha| \leq s, \partial_x^\alpha f \in L^p(\Omega)\right\}
$$
where $\alpha=\left(\alpha_1, \ldots, \alpha_d\right),|\alpha|=\alpha_1+\ldots+\alpha_d$, and the derivatives $\partial_x^\alpha f=\partial_{x_1}^{\alpha_1} \cdots \partial_{x_d}^{\alpha_d} f$ are considered in the weak sense.
$W^{s, p}(\Omega)$ is a Banach space if its norm is defined as (Sobolev norm):
$$
\|f\|_{W^{s, p}}=\sum_{|\alpha| \leq s}\left\|\partial_x^\alpha f\right\|_{L^p}
$$

In the special case $p=2, W^{s, 2}(\Omega)$ is denoted by $H^s(\Omega)$, that is, a Hilbert space for the inner product
$$
\langle f, g\rangle_{s, \Omega}=\sum_{|\alpha| \leq s}\left\langle\partial_x^\alpha f, \partial_x^\alpha g\right\rangle_{L^2(\Omega)}=\sum_{|\alpha| \leq s} \int_{\Omega} \partial_x^\alpha f \overline{\partial_x^\alpha g} d \mu
$$

\subsection{Error estimation of MLP and CNN}
\label{appx: error mlp cnn}
The fundamental neural modules of \name are MLP and CNN, thus the error estimation of \name entails the error estimation of those two modules. In this section, we cite and extend their error analysis from previous work.

\subsubsection{MLP}
Firstly, the error bound of an MLP approximating a mapping in vector spaces is well-studied in the deep learning theory. One of the existing works \cite{guhring2020error} derives the estimation based on the Sobolev regularity of the mapping. We cite the main result as the following lemma (notation modified for consistency):
    \begin{lemma}[Approximation Error of \textbf{one-dimensional MLP}~\cite{guhring2020error}]
    \label{th:1-MLP appx error}
        Let $m \in \mathbb{N}, s \in \mathbb{N}_{\ge 2}, 1 \leq q \leq \infty, M>0$, and $0 \leq n \leq 1$, then there exists a constant $C=C(m,s,q,M,n)$, with the following properties:
        For any function $f$ with m-dimensional input and one-dimensional output in subsets of the Sobolev space $W^{s, q}$:
        $$
        \norm{f}_{W^{s, q}} \le M,
        $$
        and for any $\epsilon \in (0, 1/2) $,there exists a ReLU MLP $\mathcal{N}$ such that:
        $$
        \norm{\mathcal{N}-f}_{W^{n, q}} \le \epsilon,
        $$
        and:
        \begin{eqnarray*}        
            \operatorname{size}(\mathcal{N}) &\le & C \epsilon^{-m/(s-n)} \log (\epsilon^{-s/(s-n)}), \\
            \operatorname{depth}(\mathcal{N}) &\le & C \log (\epsilon^{-s/(s-n)}).
       \end{eqnarray*}
    \end{lemma}
    However, such error bounds of one-dimensional MLP can not be directly applied to an MLP with $p$-dimensional output. A $p$-dimensional MLP can not be represented by simply stacking $p$ independent one-dimensional output networks $\{\mathcal{N}_j\}_{1\le j \le p}$ and concatenating the outputs. The key difference lies in the parameter sharing of hidden layers. To fill the gap and estimate the error of $\vb*{\beta}$, we design a special structure of MLP without parameter sharing, as explained below.

    Given $p$ independent one-dimensional output networks $\{\mathcal{N}_j: \mathbb{R}^m \rightarrow \mathbb{R}^{1}\}_{1\le j \le p}$, denote the weight matrix of the $i$-th layer of the $\mathcal{N}_j$ as $\vb*{W}_{i,j}$. The weight matrix of $i$-th layer of the $p$-dimensional MLP $\vb*{\beta}$, denoted as $\vb*{W}^{\vb*{\beta}}_{i}$, can be constructed as:
    $$\vb*{W}^{\vb*{\beta}}_{1}=\bmqty{\vb*{W}_{1,1} \\ \vb*{W}_{1,2} \\ \vdots\\  \vb*{W}_{1,p}}, \vb*{W}^{\vb*{\beta}}_{i\ge 2} = \bmqty{\vb*{W}_{i,1} & 0 & \cdots & 0 \\ 0 & \vb*{W}_{i,2}  & \cdots & 0 \\ \vdots & \vdots & \ddots & \vdots \\ 0 & 0 & \cdots & \vb*{W}_{i,p} }.
    $$
    The weight of first layer $\vb*{W}^{\vb*{\beta}}_{1}$ is a vertical concatenation of $\{\vb*{W}_{1,j}\}_{1\le j \le p}$. And the weight $\vb*{W}^{\vb*{\beta}}_{i}$ of any remaining layer $i\ge 2$ is a block diagonal matrix, with the main-diagonal blocks being $\{\vb*{W}_{i,j}\}_{1\le j \le p}$. It is easy to verify that such an approximator is computationally equivalent to the stacking of $p$ independent one-dimensional output networks $\operatorname{stack}(\{\mathcal{N}_j\}_{1\le j \le p})$. 

    Let $q=2$, $n=0$ (i.e. $\norm{\cdot}_{W^{n,q}} = \norm{\cdot}_{L^2}$), and $f:= \mathbb{R}^d \rightarrow \mathbb{R}^p$, then by Lemma \ref{th:1-MLP appx error} the approximation error is bounded by:
    \begin{align*}
    \widehat{\mathscr{E}}_{\vb*{\beta}} &= \norm{\vb*{\beta}- f}_{L^2} = \left( \sum_{j} \norm{\mathcal{N}_j- f_j}^2 \right) ^{1/2} \\
    &\le (p \varepsilon^2) ^{1/2} = \sqrt{p} \varepsilon.
    \end{align*}    
    
    And the depth and size of $\vb*{\beta}$ can be calculated by comparing with any $\mathcal{N}_j$:
    \begin{align*}
        \operatorname{depth}(\vb*{\beta}) &= \operatorname{depth}(\mathcal{N}_j) \le C \log \left(\varepsilon^{-1}\right),\\ 
        \operatorname{size}(\vb*{\beta}) &= \operatorname{size}(\vb*{W}^{\vb*{\beta}}_{1}) + \sum^{p}_{i=2}\operatorname{size}(\vb*{W}^{\vb*{\beta}}_{i}) \\
        &= p\operatorname{size}(\vb*{W}^{1,j}) + \sum^{p}_{i=2}p^2\operatorname{size}(\vb*{W}_{i,j}) \\
        & \le \sum^{p}_{i=1}p^2\operatorname{size}(\vb*{W}_{i,j}) \\
        &= p^2 \operatorname{size}(\mathcal{N}_{j}) \\
        &\le C p^2 \varepsilon^{-m / s} \log \left(\varepsilon^{-1}\right).        
    \end{align*}

    The above result can be summarized as another lemma:

    \begin{lemma}[Approximation Error of \textbf{$\mathbf{p}$-dimensional MLP}]
    \label{th:p-MLP appx error}
        Let $m \in \mathbb{N}, s \in \mathbb{N}_{\ge 2}, M>0$, then there exists a constant $C=C(m,s,M)$, with the following properties:
        For any function $f$ with $m$-dimensional input and $p$-dimensional output in subsets of the Hilbert space $H^{s}$: $\norm{f}_{H^{s}} \le M,$
        and for any $\epsilon \in (0, 1/2) $,there exists a ReLU MLP $\vb*{\beta}$ such that:
        \begin{eqnarray*}
    \norm{\vb*{\beta}-f}_{L^2} &\le& \sqrt{p} \epsilon, \\
    \operatorname{size}(\vb*{\beta}) &\leq&  C p^2\varepsilon^{-m /s}  \log \left(\varepsilon^{-1}\right), \\
    \operatorname{depth}(\vb*{\beta}) &\leq& C  \log \left(\varepsilon^{-1}\right).
    \end{eqnarray*}
    \end{lemma}

\subsubsection{CNN}
Next, we are going to derive a similar error bound for CNNs, which has not been widely studied before. A previous work \citep{zhou2020universality} derives the error bound for one-dimensional CNNs with a single channel, and the results do not apply in our case. Another related work \citep{petersen2020equivalence} established the equivalence of approximation properties between MLPs and CNNs, as the following lemma:

\begin{lemma}[Equivalence of approximation between MLPs and CNNs \citep{petersen2020equivalence}]
\label{th:equiv MLP CNN}
For any input/output channel number $N_{in},N_{out} \in \mathbb{N}_{+}$, input size $d \in \mathbb{N}_{+}$, function $f: \mathbb{R}^{N_{in} \times d} \rightarrow \mathbb{R}^{N_{out} \times d}$, $p>0$ and error $\epsilon \ge 0$:

(1) If there is an MLP $\Phi$ with $\operatorname{size}(\Phi) = S$ and $\operatorname{depth}(\Phi) = L$ satisfying $\norm{\Phi- f}_{L^p} \leq \varepsilon$, then there exists a CNN $\Psi$ with $\operatorname{size}(\Psi) \le 2S$ and $\operatorname{depth}(\Psi) = L$, and such that $\norm{\Psi- f}_{L^p} \leq d^{1/p}\varepsilon$.

(2) If there exists a CNN $\Psi$ with $\operatorname{size}(\Psi) = S$ and $\operatorname{depth}(\Psi) = L$ satisfying $\norm{\Psi- f}_{L^p} \leq \varepsilon$, then there exists an MLP $\Phi$ with $\operatorname{size}(\Phi) \le d^2S$ and $\operatorname{depth}(\Phi) = L$ and such that $\norm{\Phi- f}_{L^p} \leq \varepsilon$.
\end{lemma}
\emph{Proof idea}: The general proof idea of part (1) is to construct a lifted CNN from the MLP, such that each neuron of the MLP is replaced by a full channel of CNN, and that the lifted CNN computes an equivariant version of the MLP. And the proof of part (2) is based on the fact that convolutions are special linear maps.

The equivalence lemma (Lemma \ref{th:equiv MLP CNN}, Part (1)), combined with the approximation error of MLPs(Lemma~\ref{th:p-MLP appx error}), leads naturally to the approximation error of CNNs:

\begin{lemma}[Approximation Error of \textbf{CNN}]
\label{th:CNN appx error}
Let input/output channel number $N_{in},N_{out} \in \mathbb{N}_{+}$, input size $d \in \mathbb{N}_{+}$, $s \in \mathbb{N}_{\ge 2}$ and $M>0$, there exists a constant $C=C(N_{in},N_{out},d,s,M)$ with the following properties: for any function $f: \mathbb{R}^{N_{in} \times d} \rightarrow \mathbb{R}^{N_{out} \times d}$, $\norm{f}_{H^s} \le M$ and error $\epsilon \in (0, 1/2)$, there exists a CNN $\Psi$ such that:
 \begin{eqnarray*}
    \norm{\Psi-f}_{L^2} &\le& \sqrt{N_{out}} d \epsilon, \\
    \operatorname{size}(\Psi) &\leq&  2C (N_{out}d)^2\varepsilon^{-N_{in} d /s}  \log \left(\varepsilon^{-1}\right), \\
    \operatorname{depth}(\Psi) &\leq& C  \log \left(\varepsilon^{-1}\right).
    \end{eqnarray*}
\end{lemma}

\subsection{Proof of Theorem \ref{th:decomp approx error}: Decomposition of \name Approximation Error}

\begin{proof}
\label{pf:decomp approx error}
Firstly, extract the constant, and decompose the error by triangle inequality (subscript of norm omitted):
 \begin{eqnarray*}
     \widehat{\mathscr{E}} &=& \abs{\frac{f_{i} \circ \mathcal{G} - f_{i} \circ \mathcal{N}}{f_{i} \circ \mathcal{G}}  }  \\
     &\le & \sup_{i\in I}\qty(\frac{\operatorname{Lip}(f_{i})}{f_{i} \circ \mathcal{G}} ) \norm{\mathcal{G} - \mathcal{N}} = C\norm{\mathcal{G} - \mathcal{N}} \\
     \norm{\mathcal{G} - \mathcal{N}} & = & \norm{\mathcal{G} - \mathcal{R} \circ \mathcal{A} \circ \mathcal{E}} \\
     &= & \norm{\mathcal{G} -\mathcal{R} \circ \mathcal{P} \circ \mathcal{G} + \mathcal{R} \circ \mathcal{P} \circ \mathcal{G} - \mathcal{R} \circ \mathcal{A} \circ \mathcal{E}} \\
     &\le & \norm{\mathcal{G} -\mathcal{R} \circ \mathcal{P} \circ \mathcal{G}} + \norm{\mathcal{R} \circ \mathcal{P} \circ \mathcal{G} - \mathcal{R} \circ \mathcal{A} \circ \mathcal{E}}
 \end{eqnarray*}

The first term is exactly the reconstructor error $\widehat{\mathscr{E}}_{\mathcal{R}}$, by definition of push-forward:
\begin{eqnarray*}
    \norm{\mathcal{G} -\mathcal{R} \circ \mathcal{P} \circ \mathcal{G}}_{L^{2}(\mu)} &=&  \norm{\operatorname{Id} - \mathcal{R} \circ \mathcal{P}}_{L^{2}\left(\mathcal{G}_{\#} \mu\right)}=\widehat{\mathscr{E}}_{\mathcal{R}}
\end{eqnarray*}

And the second term is related to the approximator error $\widehat{\mathscr{E}}_{\mathcal{A}}$:
\begin{align*}
    &\norm{\mathcal{R} \circ \mathcal{P} \circ \mathcal{G} - \mathcal{R} \circ \mathcal{A} \circ \mathcal{E}}_{L^{2}(\mu)} \\
    &= \norm{\mathcal{R} \circ \mathcal{P} \circ \mathcal{G} \circ \mathcal{E}^{-1} \circ \mathcal{E} - \mathcal{R} \circ \mathcal{A} \circ \mathcal{E}}_{L^{2}(\mu)} \\
    &\leq \operatorname{Lip}(\mathcal{R}) \norm{ \mathcal{P} \circ \mathcal{G} \circ \mathcal{E}^{-1} \circ \mathcal{E} - \mathcal{A} \circ \mathcal{E}}_{L^{2}(\mu)} \\
    &= \operatorname{Lip}(\mathcal{R}) \norm{ \mathcal{P} \circ \mathcal{G} \circ \mathcal{E}^{-1} \circ \mathcal{E} - \mathcal{A} \circ \mathcal{E}}_{L^{2}\left(\mathcal{E}_{\#} \mu\right)} \\
    &= \operatorname{Lip}(\mathcal{R})\widehat{\mathscr{E}}_{\mathcal{A}}
\end{align*}
\end{proof}

\subsubsection{Extension to Non-zero Encoder Error}
\label{appx:extension}
The approximation error estimation can be naturally extended to model non-zero encoder error, as derived in \cite{lanthaler2022error}. For an encoder $\mathcal{E}$, its error $\widehat{\mathscr{E}}_{\mathcal{E}}$ is estimated by the distance to its optimal approximate inverse mapping, decoder $\mathcal{D}$, weighted by measure $\mu$. 
\begin{equation}
\begin{aligned}
\widehat{\mathscr{E}}_{\mathcal{E}} &:=& \left(\int_{I} \norm{\mathcal{D} \circ \mathcal{E}(i)-i}_I \dd \mu(i)\right)^{\frac{1}{2}} \\
\mathcal{D} &:=& \operatornamewithlimits{argmin}_{\mathcal{D}} \widehat{\mathscr{E}}_{\mathcal{E}}, \quad \text{s.t.} \quad \mathcal{E} \circ \mathcal{D}=\mathrm{Id}.
\end{aligned}
\end{equation}
Similar to reconstructor error, this error quantifies the information loss during encoding. An ideal encoder should be invertible, i.e. the decoder $\mathcal{D}=\mathcal{E}^{-1}$, thus we have $\widehat{\mathscr{E}}_{\mathcal{E}}=0$.

The estimation of $\widehat{\mathscr{E}}_{\mathcal{E}}$ depends on the specific architecture of the encoder. If the encoder can be represented by a composition of MLP and CNN, then one can derive an estimation using the lemmas in Appx.~\ref{appx: error mlp cnn}. The error of point-wise evaluation is discussed in previous work of DeepONet \citep[Sec.3.5]{lanthaler2022error} and \citep[Sec.3]{lu2021learning}.
However, the estimation of more complex encoders (e.g. the  Savitzky–Golay smoothing we used in the pushing dataset) is slightly out of the scope of this paper.

When the encoder error is non-zero, the definition of the approximator error $\widehat{\mathscr{E}}_{\mathcal{A}}$ (Eq.~\ref{eq:approximator-error}) should be modified accordingly, by replacing $\mathcal{E}^{-1}$ to $\mathcal{D}$:

\begin{equation}
\widehat{\mathscr{E}}_{\mathcal{A}}:=\left(\int_{\mathbb{R}^{m}}\|\mathcal{A}(\vb*{e})-\mathcal{P} \circ \mathcal{G} \circ \mathcal{D}(\vb*{e})\|_{\ell^{2}\left(\mathbb{R}^{p}\right)}^{2} \dd \left(\mathcal{E}_{\#} \mu\right)(\vb*{e})\right)^{\frac{1}{2}}.
\end{equation}

Also, the approximation error bound (Eq.~\ref{eq:approximation-error}) is changed to (proof similar to \ref{pf:decomp approx error}, omitted here):
\begin{equation}
    \widehat{\mathscr{E}} \leq C \qty(\operatorname{Lip}(\mathcal{G})\operatorname{Lip}(\mathcal{R}\circ\mathcal{P})\widehat{\mathscr{E}}_{\mathcal{E}}  +\operatorname{Lip}(\mathcal{R}) \widehat{\mathscr{E}}_{\mathcal{A}}+\widehat{\mathscr{E}}_{\mathcal{R}} ).
\end{equation}

\section{Environment Settings}
\label{appx: Dynamic Systems}

\begin{table}[tb!]
\centering
\caption{Typical control systems used in literature and our experiments and their dimensions.}
     
\label{tab:systems}
\begin{tabular}{@{}llcc@{}}
\toprule
System  & Description &$\vb*{u}$ \#& $\vb*{x}$ \#\\ \midrule
Pendulum & single pendulum  & 1 & 2\\ 
RobotArm & two-link robotic arm & 1 & 4\\
CartPole & one pendulum with cart  & 1 & 4\\
Quadrotor & helicopter with 4 rotors & 4 & 9\\
Rocket & 6-DoF rocket & 3 & 9\\ 
Pushing & pushing objects on various surfaces & 1 & 3\\
\bottomrule
\end{tabular}
\end{table}

\subsection{Pendulum}
\begin{eqnarray*}
& \min_{u} \int_0^{tf} \vb*{c_x}^\top(\vb*{x}(t)-\vb*{x}_{goal})^2  + c_u u^2(t)  \dd{t} \\
& \text{s.t.} \quad \dot{\vb*{x}}(t) = \bmqty{x_2(t) \\ [u(t) - m\cdot g\cdot l\sin{x_1(t)} - b\cdot x_2(t)]/I}, \\
& \vb*{x}(0) = \vb*{x}_{init} \nonumber
\end{eqnarray*}
where state $\vb*{x}=[x_1, x_2]^\top$, denoting the angle and angular velocity of the pendulum respectively, and control $u$ is the external torque. The initial condition $\vb*{x}_{init}=[0, 0]^\top$, i.e. the pendulum starts from the lowest position with zero velocity. The cost functional consists of two parts: state mismatching penalty and control function regularization, and $\vb*{c_x}=[10, 1]^\top, c_u=0.1$ are balancing coefficients. Other constants are: $m=1$, $g=10$, $l=1$, $I=1/3$.
\subsection{Robot Arm}
\begin{align*}
\vb*{M}(\vb*{x}) &= \bmqty{m_1 r_1^2 + I_1 + m_2(l_1^2 + r_2^2 + 2l_1r_2\cos(x_2)) & m_2(r_2^2 + l_1 r_2 \cos(x_2))+I_2 \\ m_2(r_2^2 + l_1 r_2 \cos(x_2))+I_2 & m_2 r_2^2 + I_2 }, \\
\vb*{C}(\vb*{x}) &= \bmqty{-m_2l_1r_2\sin(x_2)x_4 &  -m_2l_1r_2\sin(x_2)(x_3 + x_4) ) \\ m_2l_1r_2\sin(x_2) x_3 & 0 }, \\
\vb*{g}(\vb*{x}) &= \bmqty{m_1r_1 g \cos(x_1) + m_2 g (r_2  \cos(x_1+x_2) + l_1\cos(x_1)) \\ m_2 g r_2 \cos(x_1 + x_2)}, \\
\bmqty{\dot{x_1} \\ \dot{x_2}} &= \bmqty{x_3 \\ x_4},\\
\bmqty{0 \\ u}  &= \vb*{M}(\vb*{x})\bmqty{\dot{x_3} \\ \dot{x_4}} + \vb*{C}(\vb*{x})\bmqty{x_3 \\ x_4} + \vb*{g}(\vb*{x}).
\end{align*}
The RobotArm (also named Acrobot) is a planar two-link robotic arm in the vertical plane, with an actuator at the elbow.  The state is $\vb*{x}=[x_1, x_2, x_3 x_4]^\top$, where $x_1$ is the shoulder joint angle, and $x_2$ is the elbow (relative) joint angle, $x_3, x_4$ denotes their angular velocity respectively. The control $u$ is the torque at the elbow. Note that the last equation is the manipulator equation, where $\vb*{M}$ is the inertia matrix, $\vb*{C}$ captures Coriolis forces, and $\vb*{g}$ is the gravity vector. The details of the derivation can be found in \citep[Sec. 6.4]{spong2020robot}.

The initial condition $\vb*{x}_{init}=[\pi/4, \pi/2, 0, 0]^\top$, and the target state baseline $\vb*{x}_{goal}=[\pi/2, 0, 0, 0]^\top$. The cost functional coefficients are $\vb*{c_x}=[0.1, 0.1, 0.1, 0.1]^\top, c_u=0.1$. Other constants are: mass of two links $m_{1,2}=1$, gravitational acceleration $g=0$, links length $l_{1,2}=1$, distance from joint to the center of mass $r_{1,2}=0.5$, moment of inertia $I_{1,2}=1/3$.

\subsection{Cart-Pole}
\begin{equation*}
\begin{aligned}
\dot{x_1} &= x_3, \\
\dot{x_2} &= x_4, \\
\dot{x_3} &= \frac{1}{m_c+m_p \sin^2(x_2)}\left[u+m_p \sin (x_2)\left(l x_4^2+g \cos(x_2)\right)\right], \\
\dot{x_4} &= \frac{1}{l\left(m_c+m_p \sin ^2(x_2)\right)} [-u \cos(x_2)-m_p l x_4^2 \cos(x_2) \sin(x_2)-\left(m_c+m_p\right) g \sin(x_2)].
\end{aligned}
\end{equation*}

In the CartPole system, an un-actuated joint connects a pole(pendulum) to a cart that moves along a frictionless track. The pendulum is initially positioned upright on the cart, and the goal is to balance the pendulum by applying horizontal forces to the cart. The state is $\vb*{x}=[x_1, x_2, x_3 x_4]^\top$, where $x_1$ is the horizontal position of the cart, $x_2$ is the counter-clockwise angle of the pendulum, $x_3$ velocity and angular velocity of cart and pendulum respectively. We refer the reader to \citep[Sec. 3.2]{underactuated} for the derivation of the above equations.

The initial condition $\vb*{x}_{init}=[0, 0, 0, 0]^\top$, and the target state baseline $\vb*{x}_{goal}=[0, \pi, 0, 0]^\top$. The cost functional coefficients are $\vb*{c_x}=[0.1, 0.6, 0.1, 0.1]^\top, c_u=0.3$. Other constants are: mass of cart and pole $m_{c,p}=0.1$, gravitational acceleration $g=10$, pole length $l=1$.

\subsection{Quadrotor}
\label{appx:quadrotor}
\begin{align*}
    \dot{\vb*{p}} &= \vb*{v}, \\
    m\dot{\vb*{v}} &= \bmqty{0 \\ 0 \\ mg} + \vb*{R}^\top(\vb*{q})\bmqty{0 \\ 0 \\ \vb*{1}^\top \vb*{u}} ,\\
    \dot{\vb*{q}} &= \frac{1}{2} \vb*{\Omega}(\vb*{\omega}) \vb*{q}, \\
    \vb*{J} \dot{\vb*{\omega}} &= \vb*{T}\vb*{u} - \vb*{\omega} \times \vb*{J} \vb*{\omega} .
\end{align*}
This system describes the dynamics of a helicopter with four rotors. The state $\vb*{x} = [\vb*{p}^\top, \vb*{v}^\top, \vb*{\omega}^\top]^\top \in \mathbb{R}^9$ consists of three parts: position $\vb*{p}$, velocity $\vb*{v}$, and angular velocity $\vb*{\omega}$. The control $\vb*{u}\in \mathbb{R}^4$ is the thrusts of the four rotating propellers of the quadrotor. $\vb*{q} \in \mathbb{R}^4$ is the unit quaternion \citep{jia2019quaternions} representing the attitude(spacial rotation) of quadrotor w.r.t. the inertial frame. $\vb*{J}$ is the moment of inertia in the quadrotor's frame, and $\vb*{T}\vb*{u}$ is the torque applied to the quadrotor. Our setting is similar to \citep[Appx. E.1]{jin2020pontryagin}, but we exclude the quaternion from the state.

The derivation is straightforward. The first two equations are Newton's laws of motion, and the third equation is time-derivative of quaternion \citep[Appx. B]{jia2019quaternions}, and the last equation is Euler's rotation equation \citep[Sec. I.10]{truesdell1992first}. The coefficient matrices and operators used in the equations are defined as follows:
\begin{align*}
\vb*{\Omega}\left(\vb*{\omega}\right) &=\left[\begin{array}{cccc}
0 & -\omega_1 & -\omega_2 & -\omega_3 \\
\omega_1 & 0 & \omega_3 & -\omega_2 \\
\omega_2 & -\omega_3 & 0 & \omega_1 \\
\omega_3 & \omega_2 & -\omega_1 & 0
\end{array}\right],  \\
\vb*{R}(\vb*{q}) &= \bmqty{
1-2 \left(q_3^2+q_4^2\right) & 2 \left(q_2 q_3-q_4 q_1\right) & 2 \left(q_2 q_4+q_3 q_1\right) \\
2 \left(q_2 q_3+q_4 q_1\right) & 1-2 \left(q_2^2+q_4^2\right) & 2 \left(q_3 q_4-q_2 q_1\right) \\
2 \left(q_2 q_4-q_3 q_1\right) & 2 \left(q_3 q_4+q_2 q_1\right) & 1-2 \left(q_2^2+q_3^2\right) 
}, \\
\vb*{T} &= \bmqty{
0 & -l / 2 & 0 & l / 2 \\
-l / 2 & 0 & l / 2 & 0 \\
c & -c & c & -c},
\end{align*} 

We set the initial state $\vb*{x}_{init}=[[-8,-6,9]^\top, \vb*{0}, \vb*{0}]^\top$, the initial quaternion $\vb*{q}_{init} = \vb*{0}$, and the target state baseline $\vb*{x}_{goal} = \vb*{0}$. Cost functional coefficients $\vb*{c_x} = \vb*{1}$, $c_u=0.1$. Other constants are configured as: mass $m=1$, wing length $l=0.4$, moment of inertia $\vb*{J} = \vb*{1}$, z-axis torque constant $c=0.01$.

\subsection{Rocket}
\begin{align*}
    \dot{\vb*{p}} &= \vb*{v}, \\
    m\dot{\vb*{v}} &= \bmqty{mg \\ 0 \\ 0} + \vb*{R}^\top(\vb*{q})\vb*{u} ,\\
    \dot{\vb*{q}} &= \frac{1}{2} \vb*{\Omega}(\vb*{\omega}) \vb*{q}, \\
    \vb*{J} \dot{\vb*{\omega}} &= \vb*{T}\vb*{u} - \vb*{\omega} \times \vb*{J} \vb*{\omega} .
\end{align*}
The rocket system models a 6-DoF rocket in 3D space. The formulation is very close to the Quadrotor mentioned above. The state $\vb*{x}=[\vb*{p}^\top, \vb*{v}^\top, \vb*{\omega}^\top]^\top \in \mathbb{R}^9$ is same as that of Quadrotor, but the control $\vb*{u} \in \mathbb{R}^3$ is slightly different. Here $\vb*{u}$ denotes the total thrust in 3 dimensions. Accordingly, the torque $\vb*{T}\vb*{u}$ is changed to:
\begin{align*}
    \vb*{T}\vb*{u} &= \bmqty{-l/2 \\ 0 \\ 0} \times \vb*{u}
\end{align*}
We set the initial state $\vb*{x}_{init}=[10, -8, 5, -1, \vb*{0}]^\top$, the initial quatenion $\vb*{q}_{init}=[\cos(0.75),0,0,\sin(0.75)]^\top$, and the target state baseline $\vb*{x}_{goal}=\vb*{0}$. The cost functional coefficients $\vb*{c_x}=\vb*{1}$, $c_u=0.4$. Other constants are configured as: mass $m=1$, rocket length $l=1$, the moment of inertia $\vb*{J} = \operatorname{diag}([0.5, 1, 1])$

\subsection{Pushing}
\label{appx:push-data}
\begin{figure}
    \centering
    \includegraphics[width=1.0\linewidth]{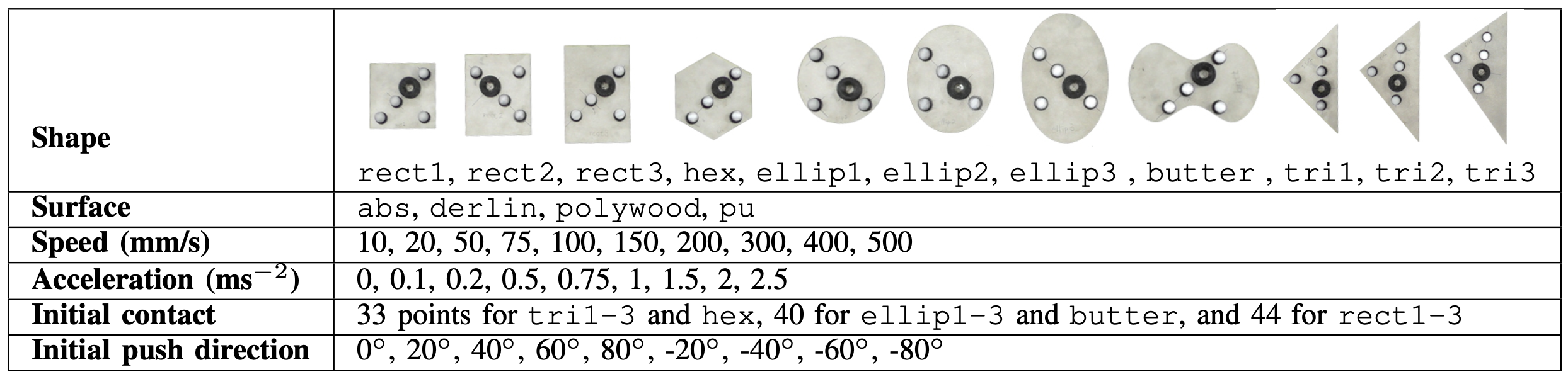}
    \caption{List of variables explored in Pushing dataset, credited to \cite{yu2016more}.}
    \label{fig:push-data-info}
\end{figure}

In this dataset, the robot executes an open-loop straight push along
a straight line of 5 cm, with different shapes of objects, materials of surface, velocity, accelerations, and contact positions and angles, see Fig. \ref{fig:push-data-info} for details.

The control and state trajectories are recorded at 250 Hz. The length of the recorded time horizon varies among samples, due to the difference in velocity and acceleration. We select acceleration $a=0.5ms^{-2}$, with initial velocity $v=0$. Then define time horizon $T=0.44s$, and extract 110 time indices per instance. 

The input to the encoder is 4167-dim, including a 768-dim gray-scale image of the shape, 3280-dim friction map matrix, 110-dim trajectory, and 9-dim other parameters (e.g. mass and moment of inertia). The input to the neural network (including CNN) is 801-dim, and the encoded vector $\vb*{e}$ is 44-dim.

\section{Details on Implementations}
\label{appx: details on exp}
For all the systems, we have 2,000 samples in each ID/OOD validation set and 100 problems in each ID/OOD benchmark set. The size of the training set varies among systems. We set 4 convolution blocks, each with 4 kernels, and the kernel length is 16. Other detailed settings are displayed in Table \ref{tab:hyper-param}. Note that the hyper-parameter settings may not be optimal, since they are not carefully tuned.

\begin{table}[tb!]
\centering
\caption{Hyper-parameter settings of the proposed \name for different systems.}
     
\label{tab:hyper-param}
\begin{tabular}{@{}lccc@{}}
\toprule
System    & \#Params & \#Train Data & \#Epochs\\ \midrule
Pendulum  & 3153 & $5 \times 10^3$ & $1 \times 10^4$ \\ 
RobotArm & 1593 & $5 \times 10^3$ & $1 \times 10^4$\\
CartPole  & 3233 & $5 \times 10^3$ & $2 \times 10^3$ \\
Quadrotor & 13732 & $1 \times 10^4$ & $5 \times 10^2$ \\
Rocket  & 10299 & $1 \times 10^4$ & $5 \times 10^2$ \\ 
Pushing  & 13174 & $1 \times 10^4$ & $2 \times 10^3$ \\ 
Brachistochrone & 3153 & $1 \times 10^4$ & $5 \times 10^3$ \\
Zermelo & 4993 & $1 \times 10^4$ & $5 \times 10^3$ \\
\bottomrule
\end{tabular}
\end{table}

PDP, DM, and synthetic control systems are implemented in CasADi \citep{Andersson2019}, which are adapted from the code repository\footnote{\scriptsize{\url{https://github.com/wanxinjin/Pontryagin-Differentiable-Programming}}}.  For classical methods, the dynamics are discretized by Euler method. To limit the running time, we set the maximum number of iterations of PDP to 2,500.

\name and other neural models are implemented in PyTorch \citep{paszke2019pytorch}. The DON number of layers is tuned in the range of $[3,5]$ for each environment, to achieve the best performance. For MLP, hyper-parameters are the same or as close as that of DON. We adjust the width of the MLP layer to reach almost the same number of parameters.  For FNO\footnote{\url{https://github.com/zongyi-li/fourier_neural_operator}}, we set the number of Fourier layers to 4 as suggested in the open-source codes, and tune the network width such that the number of parameters is in the same order as that of \name. Notice that the original FNO outputs function values at fixed time indices, which is inconsistent with our experiment setting. Thus we slightly modify it by adding time indices to its input. For GEN\footnote{\url{https://github.com/FerranAlet/graph_element_networks}}, we set 9 graph nodes uniformly spaced in time (or space) horizon, and perform 3 graph convolution steps on them. The input function initializes the node features at time index $t=0$ (multiplied by weights).  For any time index $t$, the GEN output is defined as the weighted average of all node features. Both input/output weights are softmax of negative distances between $t$ and node positions.

For fairness of running time comparison, all training/testing cases are executed on an Intel i9-10920X CPU, without GPU.

\section{More Experiment Results on Synthetic Environment}
\label{appx: more synth exp}
\begin{table}[ht]
\centering
\caption{Results of Pendulum environment.}
\label{tab:Pendulum}
\begin{tabular}{@{}lccc@{}}
\toprule
Model    & Time(sec./instance) & ID MAPE  & OOD MAPE \\ \midrule
DM & $3.80 \times 10^{-2}$ & $\diagdown$  & $\diagdown$ \\
PDP & $5.29 \times 10^{1}$ & $2.79 \times 10^{-1}$  & $1.04 \times 10^{-1}$ \\
\name & $1.80 \times 10^{-5}$ & $8.20 \times 10^{-5}$  & $2.90 \times 10^{-3}$ \\
DON & $3.36 \times 10^{-5}$ & $4.06 \times 10^{-4}$  & $1.17 \times 10^{-2}$ \\
MLP & $3.40 \times 10^{-5}$ & $2.32 \times 10^{-4}$  & $5.56 \times 10^{-3}$ \\
GEN & $6.19 \times 10^{-3}$ & $2.10 \times 10^{-3}$  & $3.43 \times 10^{-2}$ \\
FNO & $3.47 \times 10^{-4}$ & $5.94 \times 10^{-4}$  & $1.42 \times 10^{-3}$ \\
\bottomrule
\end{tabular}
\end{table}

\begin{table}[ht]
\centering
\caption{Results of RobotArm environment.}
\label{tab:RobotArm}
\begin{tabular}{@{}lccc@{}}
\toprule
Model    & Time(sec./instance) & ID MAPE  & OOD MAPE \\ \midrule
DM & $4.58 \times 10^{-2}$ & $\diagdown$  & $\diagdown$ \\
PDP & $5.62 \times 10^{1}$ & $4.73 \times 10^{-3}$  & $1.43 \times 10^{-2}$ \\
\name & $1.07 \times 10^{-5}$ & $6.11 \times 10^{-6}$  & $5.73 \times 10^{-4}$ \\
DON & $1.63 \times 10^{-5}$ & $1.28 \times 10^{-5}$  & $5.79 \times 10^{-4}$ \\
MLP & $1.56 \times 10^{-5}$ & $1.05 \times 10^{-5}$  & $1.41 \times 10^{-2}$ \\
GEN & $6.24 \times 10^{-3}$ & $2.21 \times 10^{-5}$  & $7.91 \times 10^{-4}$ \\
FNO & $3.48 \times 10^{-4}$ & $5.92 \times 10^{-6}$  & $1.42 \times 10^{-4}$ \\
\bottomrule
\end{tabular}
\end{table}

\begin{table}[ht]
\centering
\caption{Results of CartPole environment.}
\label{tab:CartPole}
\begin{tabular}{@{}lccc@{}}
\toprule
Model    & Time(sec./instance) & ID MAPE  & OOD MAPE \\ \midrule
DM & $4.63 \times 10^{-2}$ & $\diagdown$  & $\diagdown$ \\
PDP & $5.61 \times 10^{1}$ & $5.96 \times 10^{-4}$  & $4.52 \times 10^{-5}$ \\
\name & $1.69 \times 10^{-5}$ & $2.33 \times 10^{-5}$  & $6.86 \times 10^{-5}$ \\
DON & $2.53 \times 10^{-5}$ & $4.43 \times 10^{-5}$  & $4.39 \times 10^{-4}$ \\
MLP & $2.02 \times 10^{-5}$ & $4.81 \times 10^{-5}$  & $4.70 \times 10^{-4}$ \\
GEN & $6.41 \times 10^{-3}$ & $6.89 \times 10^{-5}$  & $3.39 \times 10^{-4}$ \\
FNO & $3.49 \times 10^{-4}$ & $2.48 \times 10^{-5}$  & $7.60 \times 10^{-5}$ \\
\bottomrule
\end{tabular}
\end{table}

\begin{table}[ht]
\centering
\caption{Results of Rocket environment.}
\label{tab:Rocket}
\begin{tabular}{@{}lccc@{}}
\toprule
Model    & Time(sec./instance) & ID MAPE  & OOD MAPE \\ \midrule
DM & $9.71 \times 10^{-2}$ & $\diagdown$  & $\diagdown$ \\
PDP & $7.01 \times 10^{1}$ & $1.80 \times 10^{-7}$  & $1.52 \times 10^{-7}$ \\
\name & $4.90 \times 10^{-5}$ & $8.88 \times 10^{-6}$  & $1.82 \times 10^{-4}$ \\
DON & $5.20 \times 10^{-5}$ & $1.28 \times 10^{-5}$  & $2.34 \times 10^{-4}$ \\
MLP & $2.87 \times 10^{-5}$ & $2.76 \times 10^{-5}$  & $1.60 \times 10^{-3}$ \\
GEN & $6.16 \times 10^{-3}$ & $3.57 \times 10^{-4}$  & $4.88 \times 10^{-3}$ \\
FNO & $6.33 \times 10^{-4}$ & $2.11 \times 10^{-5}$  & $2.18 \times 10^{-3}$ \\
\bottomrule
\end{tabular}
\end{table}

\newpage

\subsection{Influence of Number of Training Samples}
We test the performance of \name and other neural operators as a function of training examples on the Quadrotor environment and display it in Table \ref{tab:num-train}. As expected, the result shows that performance generally improves when the number of training samples increases. However, even when the samples are very limited (e.g. only 2000 samples), the results are still acceptable. There is a trade-off between available optimal trajectories and the generalization performance.
\begin{table}[ht]
\centering
\caption{Effect of train dataset size $N$ on Quadrotor ID MAPE }
\label{tab:num-train}
\begin{tabular}{@{}lccccc@{}}
\toprule
Model  &  2000 & 4000 & 6000 & 8000 & 10000  \\ \midrule
NASM  & $7.92 \times 10^{-6}$  & $7.27 \times 10^{-6}$ &$7.76 \times 10^{-6}$ &$8.10 \times 10^{-6}$& $6.17 \times 10^{-6}$ \\
DON  & $4.43 \times 10^{-5}$  & $4.14 \times 10^{-5}$ &$4.74 \times 10^{-5}$ &$1.89 \times 10^{-5}$& $4.08 \times 10^{-5}$ \\
MLP  & $2.40 \times 10^{-4}$  & $7.20 \times 10^{-5}$ &$9.19 \times 10^{-5}$ &$1.13 \times 10^{-4}$& $1.10 \times 10^{-4}$ \\
GEN  & $5.36 \times 10^{-4}$  & $4.25 \times 10^{-4}$ &$4.54 \times 10^{-4}$ &$5.65 \times 10^{-4}$& $6.42 \times 10^{-4}$ \\
FNO  & $8.95 \times 10^{-5}$  & $7.87 \times 10^{-5}$ &$7.02 \times 10^{-5}$ &$7.91 \times 10^{-5}$& $8.73 \times 10^{-5}$ \\
\bottomrule
\end{tabular}
\end{table}

\section{Extended Experiments on OCP with Analytical Solutions}
\label{appx: analytical ocp}
\subsection{Brachistochrone}

The Brachistochrone is the curve along which a massive point without initial speed must move frictionlessly in a uniform gravitational field $g$ so that its travel time is the shortest among all curves $y=u(x)$ connecting two fixed points $(x_1,y_1),(x_2, y_2)$. Formally, it can be formulated as the following OCP:
\begin{subequations}
\begin{align}
&&\min_{u \in C^1([x_1,x_2])} && T &= \frac{1}{\sqrt{2g}}\int_{x_1}^{x_2} \sqrt{\frac{1+\left|u^{\prime}(x)\right|^2}{y_1-u(x)}} \dd{x} \\
&&\text{s.t.}&&
u(x_1) &= y_1, \\
&& && u(x_2) &= y_2.
\end{align}
\end{subequations} 

And the optimal solution is defined by a parametric equation:
\begin{equation*}
\begin{aligned}
& x(\theta) =x_1+k(\theta-\sin \theta), \\
& u(\theta)=y_1-k(1-\cos \theta), \\
& \theta \in[0, \Theta],
\end{aligned}
\end{equation*}
where the $k, \Theta$ are determined by the boundary condition $x(\Theta)=x_2, u(\Theta)=y_2$.

We want to examine the performance of the Instance-Solution operator using the analytical optimal solution. The x-coordinate of endpoints is fixed as $x_1=0, x_2=2$, discretized uniformly into 100 intervals. Then the OCP instance is uniquely determined by $(y_1,y_2)$, which is defined as the input of neural operators. The in-distribution(ID) is $y_1^{\text{in}} \sim \mathcal{U}(2,3), y_2^{\text{in}} \sim \mathcal{U}(1,2)$ and out-of-distribution (OOD) is $y_1^{\text{out}} \sim \mathcal{U}(2.9,3.8), y_2^{\text{out}} \sim \mathcal{U}(1.9,2.8)$. The training data is still generated by the direct method(DM), not the analytical solution. The performance metric MAPE denotes the distance with the analytical solution.

The results are displayed in Table ~\ref{tab:Brachistochrone}. The \name backbone achieves the best ID MAPE and the second-best OOD MAPE, with the shortest running time. All neural operators have performance close to baseline solver DM, demonstrating the effectiveness of our Instance-Solution operator framework. It is interesting that the FNO backbone has a better OOD MAPE than DM, although it is trained by DM data. This result may indicate that the Instance-Solution operator can learn a more accurate solution from less accurate supervision data. 

\begin{table}[ht]
\centering
\caption{Results of Brachistochrone environment.}
\label{tab:Brachistochrone}
 
\begin{tabular}{@{}lccc@{}}
\toprule
Model    & Time(sec./instance) & ID MAPE  & OOD MAPE \\ \midrule
DM & $9.91 \times 10^{-2}$ & $7.33 \times 10^{-3}$  & $4.85 \times 10^{-3}$ \\
\name & $1.40 \times 10^{-5}$ & $8.05 \times 10^{-3}$  & $5.52 \times 10^{-3}$ \\
DON & $1.56 \times 10^{-5}$ & $8.31 \times 10^{-3}$  & $5.54 \times 10^{-3}$ \\
MLP & $1.49 \times 10^{-5}$ & $8.65 \times 10^{-3}$  & $6.44 \times 10^{-3}$ \\
GEN & $6.21 \times 10^{-3}$ & $9.44 \times 10^{-3}$  & $8.02 \times 10^{-2}$ \\
FNO & $3.59 \times 10^{-4}$ & $8.23 \times 10^{-3}$  & $3.82 \times 10^{-3}$ \\
\bottomrule
\end{tabular}
\end{table}

\subsection{Zermelo's Navigation Problem}
Zermelo’s navigation problem searches for the optimal trajectory and the associated guidance of a boat traveling between two given points with minimal time. Suppose the speed of the boat relative to the water is a constant $V$. And the $\beta(t)$ is the heading angle of the boat’s axis relative to the horizontal axis, which is defined as the control function. Suppose $u,v$ are the speed of currents along $x,y$ directions. Then the OCP is formalized as:

\begin{subequations}
\begin{align}
\min_{\beta \in C^1([0,T])} && T && \\ 
\text{s.t.}&& x^{\prime}(t) &=V \cos \beta(t)+u(t, x(t), y(t)), \\
&& y^{\prime}(t) &=V \sin \beta(t)+v(t, x(t), y(t)), \\
&& x(0)&=x_1, \quad y(0)=y_1, \\
&& x(T)&=x_2, \quad y(T)=y_2.
\end{align}
\end{subequations}

The analytical solution derived from PMP is an ODE, named Zermelo’s navigation formula:
\begin{equation}
\frac{\mathrm{d} \beta}{\mathrm{d} t}=\sin ^2 \beta \frac{\partial v}{\partial x}+\sin \beta \cos \beta\left(\frac{\partial u}{\partial x}-\frac{\partial v}{\partial y}\right)-\cos ^2 \beta \frac{\partial u}{\partial y}.
\end{equation}

In our experiment, we simplify $u,v$ as linear function $u(x,y)=Ax+By, v(x,y)=Cx+Dy$, where $A,B,C,D$ are parameters. And we fix the initial point $x_1=0,y_1=0$. Then an OCP instance is uniquely defined by a tuple $(x_2, y_2, V, A, B, C, D)$, which is defined as the input of our Instance-Solution operator. For the ID and OOD, we first define a base tuple $(1,1,2,0,0,0,0)$, then add with noises $\vb*{\varepsilon}^{in} \in \mathcal{U}(0,1)$ and $\vb*{\varepsilon}^{out} \in \mathcal{U}(1,2)$ for ID and OOD respectively. Again, all operators are trained with direct method(DM) solution data, and MAPE evaluates the distance toward the analytical solution.

The results are displayed in Table~\ref{tab:Zermelo}. The \name backbone has the best performance in time, ID MAPE and OOD MAPE. Most neural operators have close ID performance to DM, supporting the feasibility of the Instance-Solution operator framework.

\begin{table}[ht!]
    \centering
    \caption{Results of Zermelo environment.}
    \label{tab:Zermelo}
     
    \begin{tabular}{@{}lccc@{}}
    \toprule
    Model    & Time(sec./instance) & ID MAPE  & OOD MAPE \\ \midrule
    DM & $4.62 \times 10^{-2}$ & $1.46 \times 10^{-3}$  & $2.63 \times 10^{-3}$ \\
    \name & $1.58 \times 10^{-5}$ & $2.48 \times 10^{-3}$  & $1.31 \times 10^{-2}$ \\
    DON & $2.84 \times 10^{-5}$ & $8.67 \times 10^{-3}$  & $3.96 \times 10^{-2}$ \\
    MLP & $2.56 \times 10^{-5}$ & $6.32 \times 10^{-3}$  & $2.86 \times 10^{-2}$ \\
    GEN & $6.81 \times 10^{-3}$ & $8.64 \times 10^{-3}$  & $3.85 \times 10^{-2}$ \\
    FNO & $4.37 \times 10^{-4}$ & $2.66 \times 10^{-3}$  & $3.40 \times 10^{-2}$ \\
    \bottomrule
    \end{tabular}
\end{table}

\section{Future Direction: Enforcing the Constraints}
\label{appx: future}
In the current setting of Instance-Solution operators, boundary conditions can not be exactly enforced, since the operators are set to be purely data-driven and physics-agnostic. Instead, the operators approximately satisfy the boundary conditions by learning from the reference solution. Such a physics-agnostic setting is designed (and reasonable) for scenarios where explicit expressions of dynamics and boundary conditions are unavailable (e.g. the Pusher experiments in Section 3.2). However, our Instance-Solution operator framework may also be applied to a physics-informed setting, by changing the network backbone to the Physics-informed neural operator \cite{wang2021learning}.

And there are more important constraints in practice, such as the input constraints (e.g. maximum thrusts of quadrotor) and state constraints (e.g. quadrotor with obstacles. In our paper, they are formulated in the form of function space. Take the state $x$ for example, we define $x \in X \subseteq L(T; \mathbb{R}^{d_x})$, where $X$ is its function space, and $L(T; \mathbb{R}^{d_x})$ is the full Lebesgue space. Then $X = L(T; \mathbb{R}^{d_x})$ represents the unconstrained states, and $X \subsetneq L(T; \mathbb{R}^{d_x})$ denotes constrained states. The constrained state and control can be naturally incorporated into the PMP \cite{bettiol2021pontryagin}, thus they can be modeled in our Instance-Solution framework. Again, these constraints are satisfied approximately in our model. 

Inspired by the recent works that use SMT solver \cite{gao2013dreal} to guarantee Lyapunov stability for unknown systems \cite{chang2019neural, zhou2022neural}, we propose a possible approach to provably satisfy the OCP constraints.  Given a trained neural operator, the domain of OCP instances, and constraints, we can easily establish a first-order logic formula over real numbers, and then check the formula via the SMT solver. The solver will either prove that the neural operator satisfies all constraints for all instances within the domain or generate some counterexamples that further enhance the training dataset. If it is the latter case, the operator will be re-trained on the enhanced dataset, until the SMT solver proves that it satisfies all constraints.

\end{document}